\documentclass[12pt,preprint]{aastex}

\newcommand{\D}{$\Delta$}
\def\farcsec{\hbox{$.\!\!^{\prime\prime}$}}

\slugcomment{}

\shorttitle{The CCLQG Survey}
\shortauthors{Haberzettl et al.}

\begin{document}


\title{The Clowes-Campusano Large Quasar Group Survey: I. GALEX selected
  sample of LBGs at z$\sim$1.}


\author{L. Haberzettl}
\affil{Department of Physics and Astronomy, University of Louisville,
  Louisville, KY 40292, USA}
\email{lghabe01@louisville.edu}
\author{G.M. Williger}
\affil{Department of Physics and Astronomy, University of Louisville,
  Louisville, KY 40292, USA\\
Department of Physics and Astronomy, Johns Hopkins University, Baltimore, MD
21218, USA\\
Department of Physics, Catholic University of America, Washington DC 20064, USA}
\author{J.T. Lauroesch}
\affil{Department of Physics and Astronomy, University of Louisville,
  Louisville, KY 40292, USA}
\author{C.P. Haines}
\affil{School of Physics and Astronomy, University of Birmingham, Edgbaston, Birmingham, B15 2TT, UK}
\author{D. Valls-Gabaud}
\affil{GEPI, CNRS UMR 8111, Observatoire de Paris, 5 Place Jules Janssen, 92195 Meudon Cedex, France}
\author{K.A. Harris}
\affil{Institute for Astrophysics, University of Central
Lancashire, Preston PR1 2HE, UK}
\author{A.M. Koekemoer}
\affil{Space Telescope Science Institute, 3700 San Martin Drive, Baltimore, MD
  21218, USA}
\author{J. Loveday}
\affil{Astronomy Centre, University of Sussex, Falmer, Brighton BN1 9QH, UK}
\author{L.E. Campusano}
\affil{Departamento de Astronom\'ia, Universidad de Chile, Casilla 36-D,
  Santiago, Chile}
\author{R.G. Clowes}
\affil{Institute for Astrophysics, University of Central
Lancashire, Preston PR1 2HE, UK}
\author{R. Dav\'e}
\affil{Department of Astronomy, University of Arizona, 933N Cherry Ave.,
  Tucson, AZ 85721, USA}
\author{M.J. Graham}
\affil{CACR, California Institute of Technology, Pasadena, CA 91125,  
USA}
\author{I.K. S\"ochting}
\affil{University of Oxford, Astrophysics, Denys Wilkinson Building,
Keble Road, Oxford OX1 3RH, UK}






\begin{abstract}
The nature of galaxy structures on large scales is a key observational
prediction for current models of galaxy formation. The SDSS and 2dF galaxy
surveys have revealed a number of structures on 40-150 h$^{-1}$ Mpc scales at
low redshifts, and some even larger ones. To constrain galaxy number densities,
luminosities, and stellar populations in large structures at higher redshift,
we have investigated two sheet-like structures of galaxies at $z$=0.8 and 1.3
spanning 150 h$^{-1}$ comoving Mpc 
embedded in
large quasar groups extending over at least 200 h$^{-1}$ Mpc. We present first
results of an analysis of these sheet--like structures using two contiguous
1~deg GALEX fields (FUV and NUV) cross-correlated with optical data from the
Sloan Digital Sky Survey (SDSS). We derive a sample of 462 Lyman Break Galaxy
(LBG) candidates coincident with the sheets. Using the GALEX and SDSS data, we
show that the overall average spectral energy distribution of a LBG galaxy at
$z \sim$1 is flat (in f$_{\lambda}$) in the rest frame wavelength range from
1500\AA\, to 4000\AA, implying evolved populations of stars in the LBGs.
From the luminosity functions we get indications for overdensities in the two
LQGs compared to their foreground regions. Similar conclusions
come from the calculation of the 2-point correlation function, showing a
2$\sigma$ overdensity for the LBGs in the $z \sim 0.8$ LQG on scales of 1.6 to
4.8~Mpc, indicating similar correlation scales for our LBG sample as
their $z\sim 3$ counterparts. 
\end{abstract}





\section{Introduction}

The relation between galaxy populations, environment, and the quasar/AGN phase
of galaxy evolution is crucial for a complete picture of galaxy and structure
formation. Galaxy redshift surveys have revealed the cosmic web, a cellular
distribution  \citep[e.g.][]{2001MNRAS.328...79D} on scales of
40-150~h$^{-1}$ Mpc or more. 
In particular, large structures were found in the SDSS and 2dF redshift survey
\citep[][]{2006ApJS..162...38A,2004MNRAS.349.1397C}, tracing galaxy populations
and star formation in the cosmic web. The $\Lambda$CDM model
predicts weakly non-linear structures of these dimensions at low redshift,
but simulations \citep[e.g.][]{2002ApJ...573....7E} indicate that ``Great
Wall''-like sheets become rare at $z \sim$1. Clearly the detection of very
large filaments and the presence of a large number of high redshift clusters
are an interesting test of current cosmologies.
Distinct blue and red galaxy sequences at $0<z<0.7$ indicate significant
environmental effects on the red fraction at fixed luminosity
\citep[e.g.][]{2004ApJ...615L.101B,2007ApJS..172..270C}. Environmental effects
are especially key in 
triggering/quenching star formation through mergers, harassment, and gas
stripping \citep[][]{2005ApJ...623..721P}, though on cluster outskirts, some
star formation also goes on
\citep[][]{2002A&A...382...60D,2005A&A...430...59C}. There is clear evidence that galaxies in filaments falling
into clusters of galaxies undergo bursts of star formation prior to reaching the
cluster, both at low
redshifts \citep[][]{2008MNRAS.388.1152P} and at $z \sim 1$ \citep[][]{2008MNRAS.391.1758K}
even though the specific star formation rates may be lower for those
galaxies in the red sequence \citep[e.g.,][]{2002aprm.conf..283N}. Understanding
these mechanisms is essential to assess the role of environment in galaxy
evolution.

Extremely dense regions should show the most extreme environmental effects.
The epoch $z\sim 1$ is a key point, where the galaxy luminosity function (LF)
still can be readily probed to faint levels and environmental effects appear
to affect galaxy evolution strongly. Recent work
\citep[][]{2007MNRAS.376.1425G,2007ApJ...660L..43N} indicates that $z\sim
1-1.5$ is where the galaxy color-density 
relation establishes itself: at lower $z$ red galaxies prefer dense
environments, while at earlier times star-forming and passive galaxies inhabit
similar environments, with possible evidence for an inversion
\citep[][]{2007MNRAS.tmp.1119C,2008MNRAS.383.1058C,2007A&A...468...33E}. 
However, \citet[][]{2007ApJ...654..138Q} find 
that the color-density relation (in the sense of redder galaxies being located
in denser environments) extends to redshifts $z > 2$. Hence studies of galaxies
in overdense environments at this epoch promise to yield crucial insights into
this critical part of galaxy formation theory. 

Quasars may signal gas-rich merger environments
\citep[][]{2008ApJS..175..356H}, and large quasar groups (LQGs) are
potentially unique structure markers on scales up to hundreds of Mpc.  LQGs
could therefore provide a very efficient means to study both quasars and
galaxies in a wide variety of environments, from low to high densities.
Analogous to star formation quenching \citep[e.g.,][]{2005A&A...430...59C},
quasars form preferentially in cluster outskirts at $z\sim 0.4$, and delineate
the underlying large-scale structure
\citep[e.g.,][]{2002MNRAS.331..569S,2004MNRAS.347.1241S}. Although LQGs are too
large to be virialized at their redshifts, they are still highly biased
tracers of what may be the largest scale density perturbations. The average
LQG space density is 7~Gpc$^{-3}$ at $0.3<z<1.9$
\citep[][]{2007ARep...51..820P}, or $\sim 3-4\times$ below supercluster number
densities at $z<0.1$ \citep[$\sim
25$~Gpc$^{-3}$,][]{2007MNRAS.379.1343S}. They form two classes, $\sim 70$\%
with 6-8 members, average sizes of 90~h$^{-1}$ Mpc, and overdensities of $\sim
10$, and $\sim 30$\% with 15-19 member, average scales of 200~h$^{-1}$ Mpc
and overdensities of 4. Six
such mega-structures were found by \citet[][]{2007ARep...51..820P} in the 2dF
quasar survey (750~deg$^2$), implying 500-1000 ``jumbo'' LQGs in the sky.

One aspect to understanding the formation of Large Scale Structure (LSS) is
the star formation within them and its connection to the
environment. Lyman Break Galaxies \citep[LBGs;][]{1996ApJ...462L..17S}
are one representative class of galaxies for ongoing star formation, forming
stars on relative high levels \citep[][]{1998ApJ...498..106M}. The Lyman break
at 912\,\AA\, (rest frame) is a spectral signature which makes it relatively
easy to detect large numbers of LBGs at high redshifts (z$>$2) by color
selection, using the drop-out technique
\citep[][]{1995AJ....110.2519S}. Therefore LBGs   
were one of the first confirmed population of high redshift galaxies
\citep[][]{1995AJ....110.2519S}. 
They were found to be more metal rich than expected, with metallicities
Z$>$0.1Z$_\odot$ \citep[][]{2000ApJ...533L..65T,2001ApJ...554..981P}. Their
stellar populations are similar to those found in local star-burst galaxies
with stellar masses of several 10$^9\,M_\odot$ up to 10$^{10}\,M_\odot$ for
L$^*$ luminous LBGs \citep[][]{2001ApJ...559..620P}. Therefore, they show only
0.1 times the mass of present day L$^*$ galaxies. LBGs at z$>$2 are expected
to be the 
precursors for the present day massive galaxies evolving via mergers into
massive elliptical galaxies at $z$=0 \citep[][]{2002ApJ...564...73N}. However,
the $z\sim1$ LBGs are likely to be the progenitors of rather less massive
ellipticals, $0.1-1 L^*$, since there is only little growth in the elliptical
galaxy population after $z\sim1$. 
Although the bulk of the ongoing star formation at 2$\le$z$\le$4 can be
observed in the optical wavelength range, dust plays a substantial role in
high redshift galaxy evolution. In contrast to early assumptions as by
\citet[][]{1998ApJ...498..106M}, studies of the dust attenuation of
\citet[][]{2003ApJ...587..533V} showed that LBGs are affected by extinction up
to $\sim$5 mag (rest frame 1600~\AA), requiring significant corrections to
star formation rates (SFRs) from LBGs.  

Other types of galaxies such as luminous and ultra luminous infrared galaxies
(LIRGS and ULIRGS) require identification in the mid-IR. In those wavebands
the observing capabilities (e.g., Spitzer) mostly do not allow coverage of
large areas with high spatial resolution down to low sensitivities. LBGs are
easier to survey and therefore offer an efficient statistical measure of star
formation in the early Universe, as long as results are interpreted with
attention to extinction from dust. 

LBGs reflect value as mass tracers by virtue of their correlation
properties. At $z$\,$^>_\sim$\,2 LBGs show strong clustering with power law
slopes of the angular correlation function of 
$\beta\,\sim$\,0.5-0.8 on scales of 30$^{\prime\prime}$ to 100$^{\prime\prime}$
\citep[][]{2002ApJ...565...24P,2003A&A...409..835F,2007A&A...462..865H}, with
brighter LBGs clustering more strongly. Their correlation lengths are
$\sim$4-6\,h$^{-1}$  Mpc.  A similar correlation length range was
observed between AGN and LBGs at $z\sim 3$ \citep[][]{2005ApJ...627L...1A},
for AGN black hole masses of $5.8< \log(M_{BH}/M_\odot) < 10.5$.

Although we have very good knowledge about LBGs at $z$\,$^>_\sim$\,2, we are
only now observing $z\sim$1 LBGs in large numbers. First results for LBGs at
this critical epoch of galaxy evolution were published by 
\citet[][]{2006A&A...450...69B,2007MNRAS.380..986B} using CDF-South data. They
defined their sample of z$\sim$1 LBGs using GALEX observations combined with
multi-wavelength coverage from X-ray to mid-IR. The majority of their LBG sample
consists of disk-dominated galaxies with a small number ($\sim20$\%) of
interacting/merging members and almost no spheroidals. They found that
UV-luminous LBGs are less affected by dust than UV-faint
ones. \citet[][]{2007MNRAS.380..986B} also showed from comparison to model
spectral energy distributions (SEDs) that the averaged spectra of LBGs
indicate luminosity-weighted (SEDs which are dominated by the brightest stars
in the galaxy) ages between 250 and 500 Myr.  
LBGs at $z\sim 1$ therefore are an important contributor to the UV luminosity
density and represent the majority of star formation, as disk and irregular
galaxies identified in the LBG sample of
\citet[][]{2006A&A...450...69B,2007MNRAS.380..986B} should represent the
majority of the star formation at $z\sim 1$ based on UV to mid-IR SEDs from GEMS
\citep[][]{2005ApJ...630..771W, 2005ApJ...625...23B}.

In contrast to Burgarella et al., we report first results from a
study of star forming galaxies at $z\sim$1 in regions with high quasar
  overdensities, as opposed to the presumed typical CDF-S. The high quasar
space density enables us to make a comparison to the higher redshift LBG-AGN
correlation of \citet[][]{2005ApJ...627L...1A}. We describe our data and
selection method in \S~\ref{observ} followed by a description of our photo-z
determinations in \S~\ref{redshifts}. Results from the analysis of the
stacked SEDs of subsamples of our LBGs in different redshift intervals are
presented in \S~\ref{sedfit}+\ref{lbgsamp}, followed by summaries and
conclusions in \S~\ref{sumcon}. Our sample increases the number of published
LBG galaxies at $z\sim$1 by about a factor of two. Throughout the paper we use a
cosmology with $H_0\,=\,71$\,km\,s$^{-1}$\,Mpc$^{-1}$, $\Omega_M\,=\,0.3$ and
$\Omega_\Lambda\,=\,0.7$. All distances quoted are comoving distances, unless
otherwise stated.  

\section{Observations}
\label{observ}
\subsection{Target Field}

\begin{figure*}
\epsscale{1}
\plotone{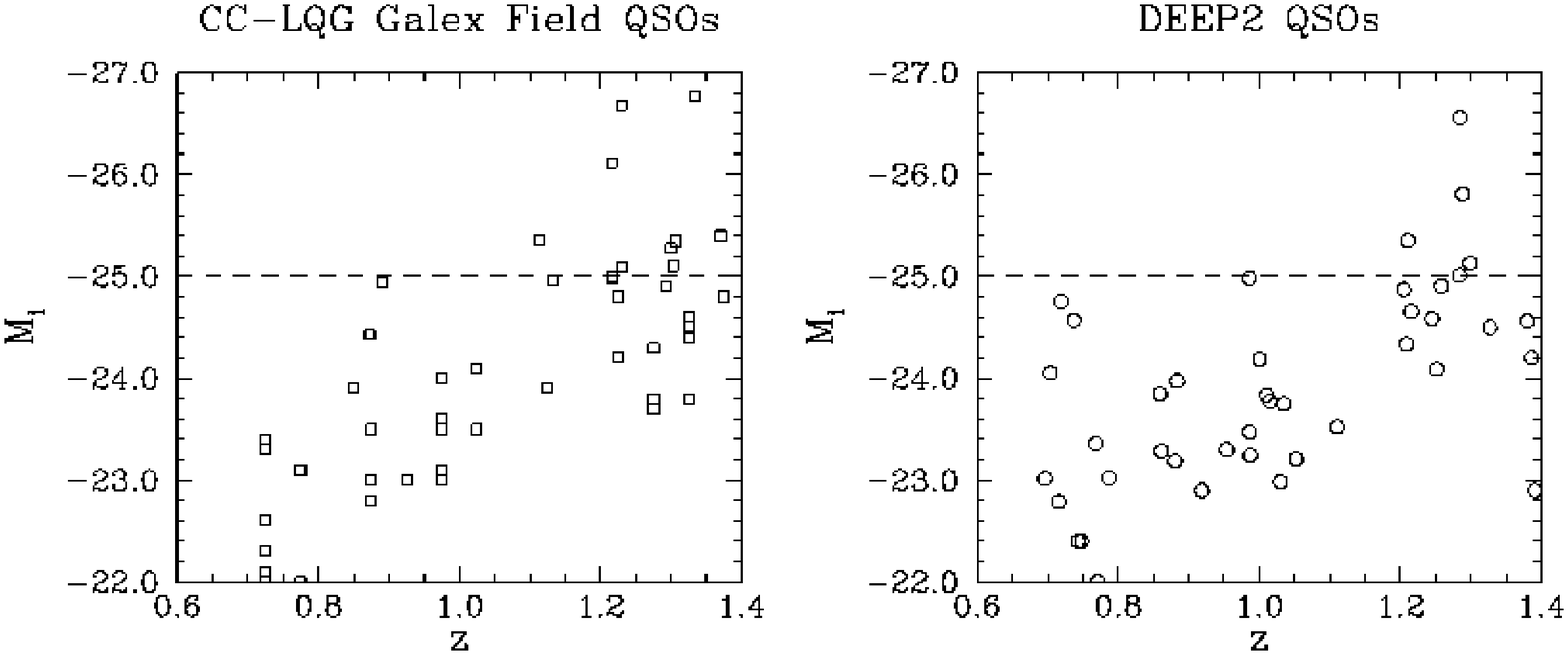}
\caption{The distribution of QSO and AGN M$_{\rm I}$ magnitudes versus $z$ for
  our two {\it GALEX} fields (left, 2.2 deg$^2$) and the four DEEP2 fields
(right, 3 deg$^2$; Coil {\it et al}.\ 2007).  The QSO redshifts for the
{\it GALEX} fields are from \citet[][]{1991MNRAS.249..218C,1994MNRAS.266..317C}
and \citet[][]{1999PhDT..........N}. Note that this plot does not include the
objects 
from the photometric selected catalog of \citet[][]{2007AAS...21114202R},
which includes another $\sim$40 QSO and AGN candidates to g$\sim$22.0 in
the {\it GALEX} fields.
}
\label{qsodense}
\end{figure*}

\placefigure{qsodense}
We observed part of the Clowes-Campusano LQG (CCLQG), which is the
largest known LQG 
and has the most members. It contains at least 18 bright quasars at
1.2$<z<$1.5 and a spatial overdensity of 3 for B$_J\,\le\,20.0$\,mag 
\citep[][]{1991MNRAS.249..218C,1994MNRAS.266..317C,1995MNRAS.275..790G,1999MNRAS.309...48C,2007AJ....134..102S}.
The CCLQG covers $\sim$2.5\,deg\,x\,5\,deg ($\sim$120x240 h$^{-2}$ Mpc$^2$
) and is 590  h$^{-1}$ Mpc deep. It was discovered in a
$\sim$\,25.3\,deg$^2$ 
objective-prism survey using UK Schmidt plate data \citep[plate UJ5846P or
ESO/SERC field 927,][]{1991MNRAS.249..218C} and represents one of the largest
known structures at $z>$1. It is 3$\times$ denser in bright
(M$_I<-25$) quasars compared to the DEEP2 fields
(Fig.~\ref{qsodense}). Previous studies of the CCLQG showed an
associated factor 3 overdensity of Mg{\sc ii} absorbers at 1.2$<z<$1.5
\citep[][]{2002ApJ...578..708W}, which are linked to luminous galaxies
\citep[e.g.][]{1997ApJ...480..568S,1997A&A...328..499G}.
There is also a foreground $z\sim 0.8$ LQG containing $\geq 14$ quasars and
spanning $3.5^\circ \times 3^\circ$ on the sky. Studies of the galaxy
populations in the LQGs showed $\sim$30\% overdensities of red galaxies
(I-K$>$3.4) at $z$=0.8 and z=1.2
\citep[][]{2001MNRAS.323..688H,2004A&A...421..157H}. The galaxy colors are 
consistent with an evolved population, and form sheets which span a
$\sim$40$^\prime\times$34$^\prime$ subfield which we imaged deeply in VI using
the CTIO Blanco 4m telescope
\citep[][]{2001MNRAS.323..688H,2004A&A...421..157H}. Smaller
5$^\prime\times$5$^\prime$ SOFI 
subfields of near-IR imaging reveal three clusters at $z$\,$\sim$\,0.8 and a
pair of merging clusters at $z$\,=\,1.2 associated with a CCLQG member quasar
\citep[][]{2004A&A...421..157H}.

\subsection{Data}

\placetable{exposuretimes}
For this study we have imaged two slightly overlapping 1.2~degree fields 
within the CCLQG in the Far-UV (FUV; $\lambda_{eff} = 1538.6$\AA) and Near-UV
(NUV; $\lambda_{eff} = 2315.7$\AA) filter bands, using the UV satellite GALEX
(GALaxy Evolution eXplorer). The observations were part of the Guest
Investigator program, cycle 1 proposal 35. The data used in this
  work consist of two $\sim$ 20000\,s
exposures in the FUV and two $\sim$\,35000\,s exposures in the NUV filter band
(Tab.~\ref{exposuretimes}). The pipeline reduction was done by the GALEX team,
including the photometric calibration. We are able to detect point sources with
SExtractor v.2.5.0 \citep[][]{1996A&AS..117..393B} to m$_{NUV,FUV}$ $\sim$25.5
mag. For further analysis we used the MAG\_ISO parameter as measured by
SExtractor, which gives total magnitude for the measured objects. For details
see the SExtractor
manual\footnote{{\tt http://terapix.iap.fr/rubrique.php?id\_rubrique=91}}. A
detailed description of the completeness analysis of the 
data can be found in \S~\ref{complet}. Optical complementary data are from the
Sloan Digital Sky Survey DR5 \citep[June, 2006;][]{2007ApJS..172..634A}, which
is sensitive to limiting 
magnitudes of u$ = 22.0$, g$ = 22.2$, r$ = 22.2$, i$ = 21.3$, $z$$ = 20.5$. With
the SDSS data, we have 7 band photometric information available for most
of our UV-selected sample (see \S~\ref{sampsel}).
From the SDSS we obtained model magnitudes in the five filter bands, flux
values, and spectroscopic and photometric redshift information. Since
confusion of sources represents a 
significant effect in the GALEX data, we chose from the SDSS list the nearest
primary object (as defined by the SDSS) within 4\farcsec5 to match our UV
detections. 
       
\subsection{Sample Selection}
\label{sampsel}

\begin{figure}
\epsscale{1}
\plotone{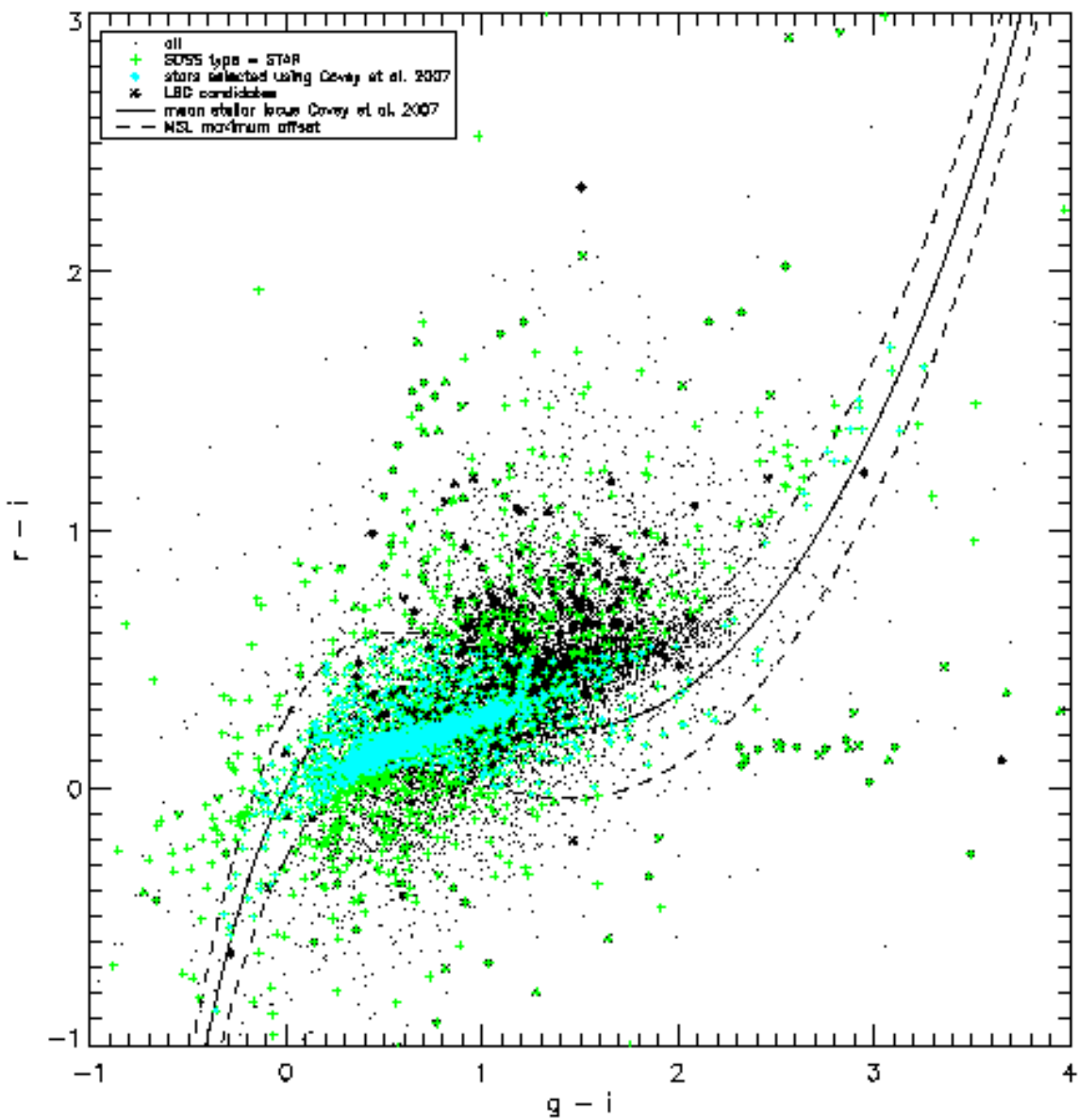}
\caption{Color-color diagram showing the location of the mean stellar locus
  (MSL, red solid line) as defined by \citet[][]{2007AJ....134.2398C}. The
  blue dashed lines show the maximum offset to the analytical solution. Point
  sources within the maximum offset to the MSL are considered stars (cyan
  dots, making the streak below the MSL). Point sources outside the
  maximum offset to the MSL are selected as galaxy candidates (green
  dots). The red stars indicate the location of our LBG candidate sample.}
\label{starsel}
\end{figure}

\placefigure{starsel}
The analysis described in the following parts of the paper are based on a
sample selected in the two UV filter bands of the GALEX data. The source
catalog has been created using SExtractor v.2.5.0.
The complete catalog consists of 15688 sources applying a detection threshold
of 3.0 $\sigma$ and an
analysis threshold of 1.5 $\sigma$, using the default convolution filter of
SExtractor. Additionally, we also used weight maps as well as flag images to
exclude bad regions at the edge of the GALEX images and saturated sources
within our SExtractor search. The weight maps and flag images were constructed
using {\sc weight watcher version 1.7} \citep[][]{2008ASPC..394..619M}. A
cross-correlation with the SDSS DR5 
resulted in 14316 sources (matching radius: 4\farcsec5). To clean our sample
from false detections 
(e.g. bright star contaminations, reflections) we only selected objects which
have a SExtractor extraction {\it FLAGS $\le\,2$} in the NUV filter, which
resulted in a subsample of 13760 objects (final UV selected sample). 
For the star-galaxy discrimination
in the sample we use the {\it PhotoType values} of the SDSS DR5, which works on
the 95\% level for object with r$\le22.2$\,mag
\citep[][]{2007ApJS..172..634A}. We selected
all objects which were marked as {\it GALAXY} in the SDSS data, yielding
10982 galaxies. However, since the PSF of the GALEX data is relatively poor
($\sim$\,4\farcsec5 for the FUV and $\sim$\,6\arcsec for the NUV filter), it is
difficult to distinguish between real point sources (e.g. stars or quasars)
and higher redshift extended sources which are point-like in the GALEX
beam. Therefore, to reduce the contamination of our LBG sample with faint
stars, we followed our more conservative approach. 
To account for smaller point like objects we also selected all
objects which are marked as {\it STAR} and are located outside the mean
stellar locus (MSL) as defined by \citet[][]{2007AJ....134.2398C}. For this
selection we used the analytic fit for the MSL of 
\citet[][]{2007AJ....134.2398C} in the g-i vs. r-i color-color space (see
Fig.~\ref{starsel}) and included all objects in our sample which have an
offset to the MSL larger than the maximum offset indicated by the blue dashed
lines. With these additional selection criteria, we ended up with a
final sample of 11635 galaxy candidates with a photometric redshift
distribution as shown in Figure~\ref{redhisto} (red bars). 

To identify $z\sim 0.5-1$ LBGs, we applied a FUV dropout technique, using the
selection criteria of \citet[][]{2006A&A...450...69B} and including only
objects classified as {\it GALAXY} by the SDSS DR5 or outside the MSL and with
m$_{NUV}<23.5$\,mag and m$_{FUV}-$m$_{NUV}>2$\,mag. This resulted in a final
sample of 1263 LBG galaxy candidates (618\,deg$^{-2}$), which is only about
half the detection rate of \citeauthor[][]{2006A&A...450...69B}, who found
1180 LBG candidates per deg$^2$ in the CDF-S. Without these additional selection
parameters our selection would have resulted in a sample of 2566 LBG
candidates or 1256 deg$^{-2}$, which is comparable to the results of
\citeauthor[][]{2006A&A...450...69B}. However, unlike
\citeauthor[][]{2006A&A...450...69B}, we only have 7 band GALEX+SDSS photometry
  as opposed to their much wider UV to mid-IR wavelength coverage and higher
  resolution imagery. Therefore, to reduce the contamination of our
LBG sample with faint stars, we followed our more conservative approach.

\subsection{Completeness and Confusion}
\label{complet}

\begin{figure*}
\epsscale{1.1}
\plottwo{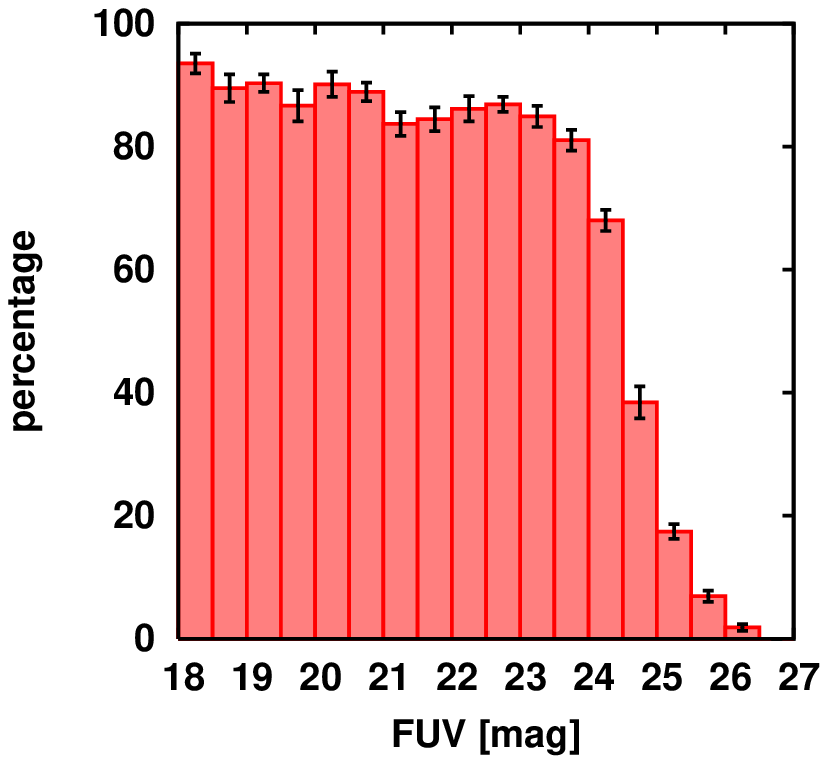}{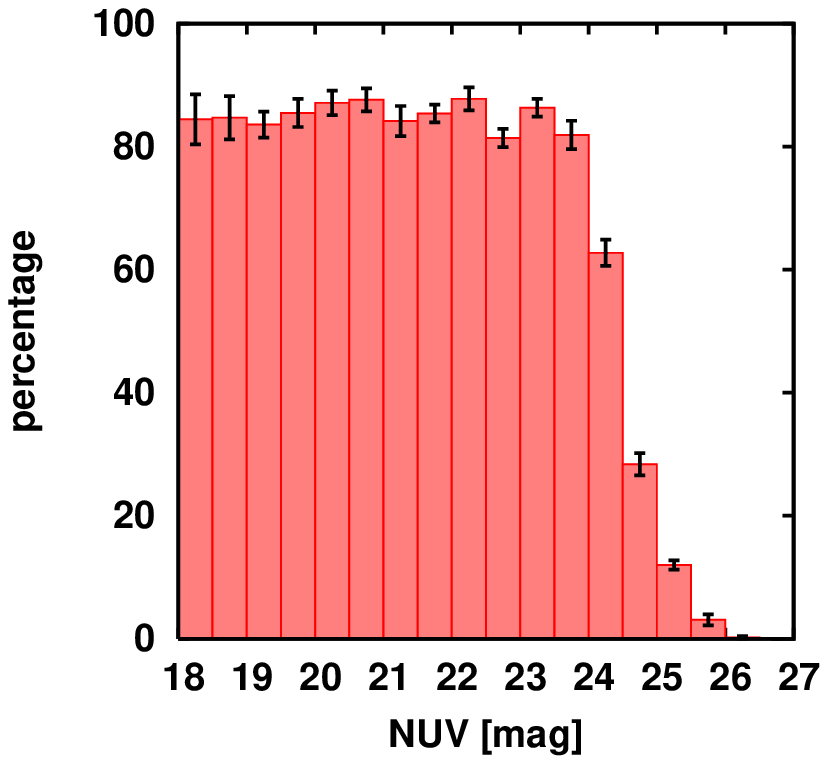}
\plottwo{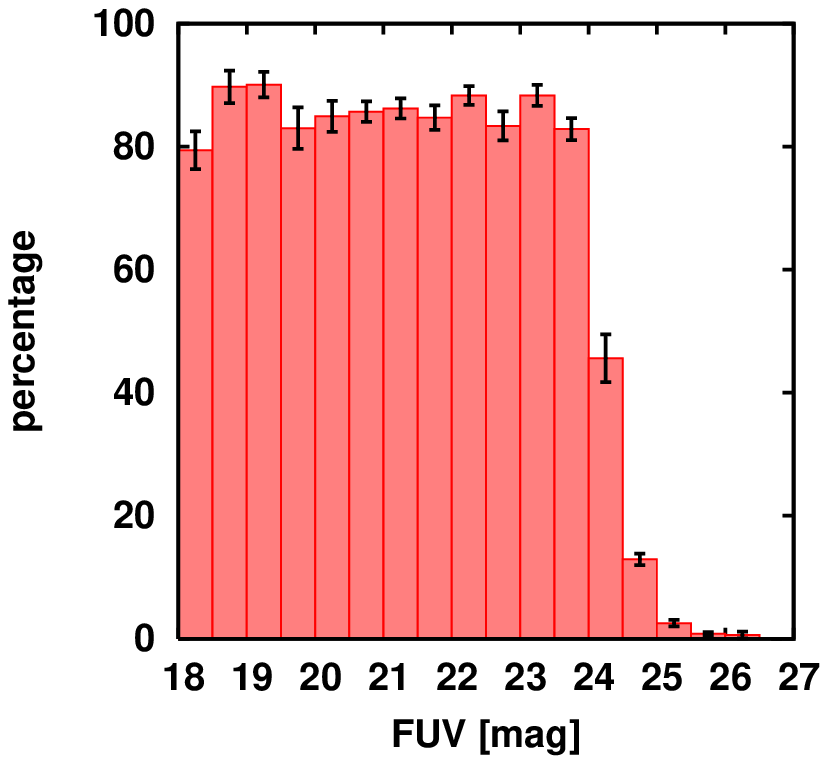}{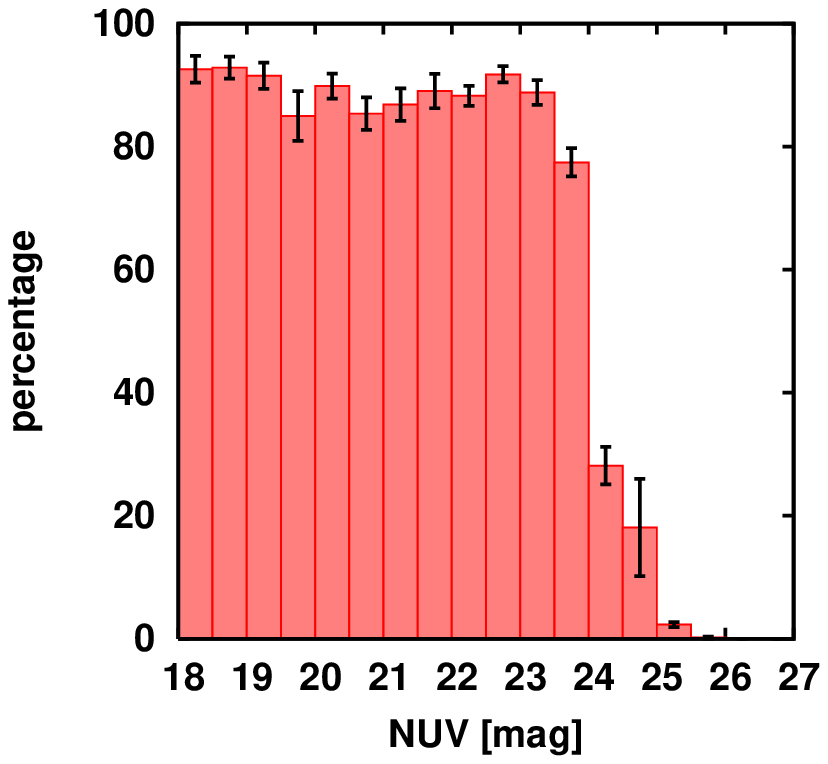}
\caption{Results of the completeness simulations for the northern (upper row)
  and southern field (lower row) in the FUV (left column) and NUV (right
  column) filters. The marginally low values for the detection efficiency at
  AB$<$24.0 can be explained by confusion due to the large GALEX PSF.
}
\label{completness}
\end{figure*}

\placefigure{completness}
We used a Monte Carlo-like approach to check the completeness of galaxy counts
in the FUV and NUV filter bands. The process relies on a well-known artificial
sample of galaxies. To keep the basic image information (e.g. noise
pattern, pixel size, PSF) of the original data, we simulated the
artificial 
galaxy sample as follows. We first removed all sources detected by SExtractor
from the images, to get a source-free image with the real observed
background. This was done using the SExtractor check-image ({\it
  CHECKIMAGE\_TYPE = -OBJECTS}). In the cleaned images we
then placed a list of simulated galaxies constructed by running the {\it IRAF}
task {\it gallist}.  We created a list of 2000 synthesized 
galaxies randomly placed in a $\sim$0.7 deg$^2$ field. To simulate the
artificial galaxy sample, we chose a luminosity distribution based on a
power law with an exponent of 0.1. The simulated galaxy list consists of objects
with total magnitudes between 14 and 25.5 mag and redshifts out to
$z$\,=\,1.3. To create the galaxies in the source-free GALEX images, we
applied 
the {\it IRAF} task {\it mkobjects}. This procedure resulted in a
well-defined data set with a known galaxy sample which we could use for
further analysis. We then photometrically analyzed the artificial galaxies and
derived their total magnitudes using the galaxy fitting routine {\it
  galfit} \citep[][]{2002AJ....124..266P} on the simulated data. In the final
step, we searched the images for galaxies with
SExtractor by applying the same extraction parameters as for our science data
(see \S~\ref{sampsel}), and compared the resulting list with the input list 
to compute detection efficiencies. We repeated this procedure ten times and
calculated the mean and standard deviation for the detection efficiencies (see
Fig.~\ref{completness}). For all four cases (FUV, NUV data for the northern and
southern fields) our detection efficiency is around 80-90\% for objects
with total magnitudes down to 24\,AB. The detection efficiency decreases to
$\sim$60\% at 24.5\,ABmag for the northern FUV and NUV images and 45-30\,\%
for the southern images. For objects with total magnitudes
fainter 24.5, the detection efficiency drops below 40\% for the northern
and below 20\% for the southern field. The marginally low efficiencies (only
80-90\%) for 
the bins brighter than 24\,AB can be explained by blending effects due to
confusion resulting from the large pixel size of GALEX (1\farcsec5) and the
large PSF in the FUV ($\sim$5\arcsec) and NUV filters ($\sim$\,6\farcsec7). 

We therefore checked the GALEX detections for multiple
counterparts using higher resolution optical images obtained at the CFHT with
Megacam. 
The images reach limiting magnitudes of r=27 and z=25, with a mean image
quality of $\sim$1~arcsec in both filters \citep[for more details
see][]{haberzettl09}. Confusion can effect the photometry either by
  increasing the flux of detected objects directly or by changing the
  background estimates in the surrounding of the detected objects if the
  backround is estimated locally. Since we derive a global estimate of the
  background within SExtractor by using a mesh size of 64 pixels
  (significantly larger than the PSF), this effect
  should be small and can be neglected. More important is the change in
  photometry from additional objects within one FWHM. We therefore
  matched all sources in the NUV-, r-, and z-band 
  using a radius of 3$^{\prime\prime}$ diameter, which covers a slightly
  larger area than the 5\farcs4 FWHM of the GALEX NUV
  data. Our analysis show that 19\% of the complete sample
has 2 or more counterparts to r$\sim$27. Restricting our study to
counterparts with r$\le 24.5$, the multiple counterpart fraction
decreases to 16\%. For the LBG candidates with NUV$\le23.5$, we found 2 or more
counterparts for 22\% of the sample (r$\le$24). Although these values
for confusion are slightly higher then previously reported
\citep[e.g. 13\%][]{2007ApJS..173..659B}, we do not correct our sample using 
deconvolution methods, since our results are based on averages over samples of
galaxies. The effects of confusion will be minor, compared to the scatter in
our SEDs (details in \S\ref{sedfit}-\ref{lbgsamp}). To this point, we did not
restrict optical counterpart colors. If we restrict optical counterparts to
r-z$\le$0.5 (or f$_z\ge$1.5~f$_r$) which means the objects are likely to
  be either relatively old ($\ga$5 Gyr) or dust reddened and in both cases have
  significantly reduced UV-fluxes, the fraction with $\ge$2 counterparts
reduces to 16\% for the complete sample (NUV$\le$23.5) and 10~\% for the LBG
subsample.  

\begin{figure}
\epsscale{1}
\plotone{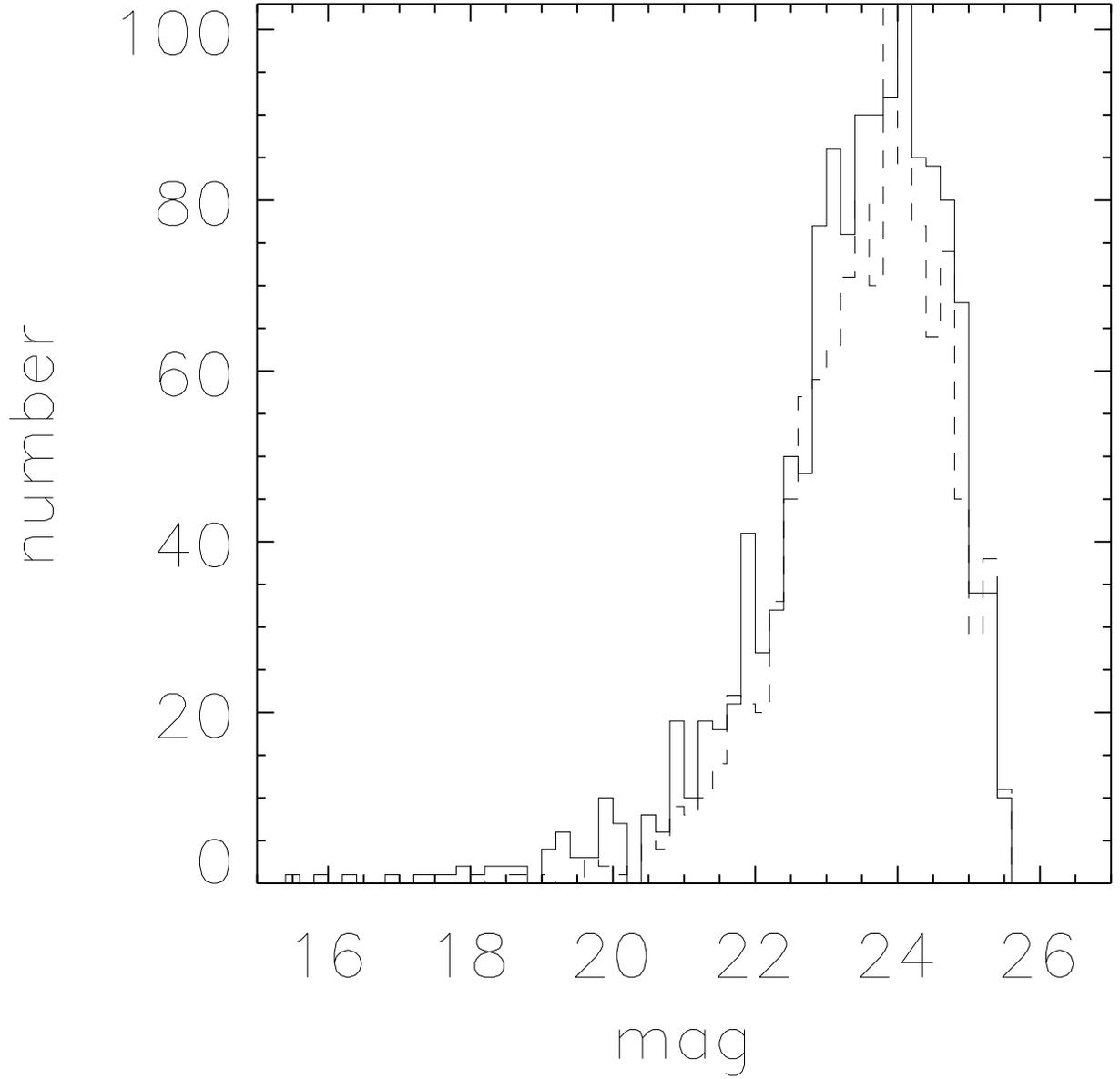}
\caption{Comparison of the distribution in magnitudes for real (dashed)
    and simulated (solid) NUV GALEX data. The distributions are consistent
    with each other. Poisson errors in the real data are omitted for
    clarity. See text for simulation details.}
\label{simuhist}
\end{figure}

\placefigure{simuhist}

Additional information about confusion results from the number of sources
per beam (beams per source). We followed the approch by
\citet[][]{2001AJ....121.1207H} as used by
\citet[][]{2007MNRAS.380..986B}. Restricting the galaxy sample to
NUV~$\le$~23.5 results in (s/b)$^{3\sigma}_{conf}$~=~0.0038 sources per beam
(265 beams per source). This is much shallower than the 3$\sigma$ confusion
limit of (s/b)$^{3\sigma}_{conf}$~=~0.063 or 16 beams per source reported by
\citet[][]{2007MNRAS.380..986B} indicating that confusion will not effect
our results significantly. 
\begin{eqnarray}
(s/b)^{3\sigma}_{conf}~=~\frac{N_{sources}}{\Omega_{GALEX}/\Omega_{beam}}
\label{confeqn}
\end{eqnarray}

\begin{figure*}
\epsscale{1}
\plotone{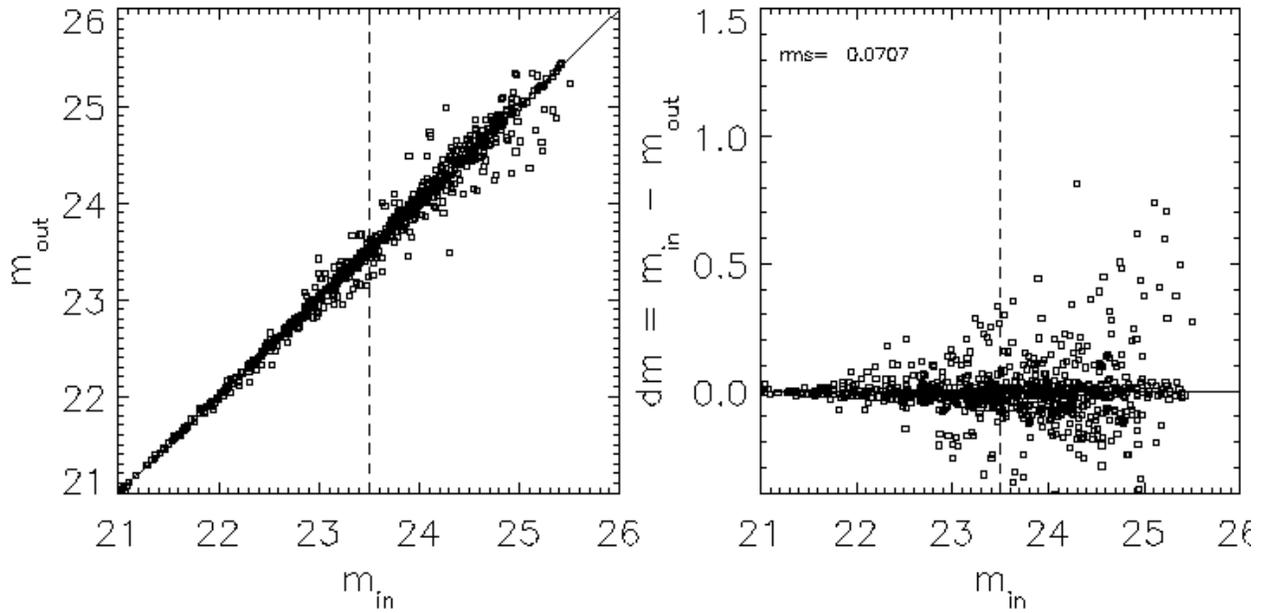}
\caption{Effect of confusion on photometric results of NUV GALEX
    data. Comparison of magnitudes before (m$_{in}$) and after placing
    additional artificial galaxies in the field (m$_{out}$). For galaxies with
    m$_{in} \le$23.5 (reflecting the selection criteria used for the LBG
    candidate sample) the change in magnitude is relatively small with a rms
    for the offset dm$_{rms}$ = 0.0707.}
\label{simuconf}
\end{figure*}

\placefigure{simuconf}

To test the effect of confusion on our photometry, we simulated GALEX NUV
images using SkyMaker
3.1.0\footnote{http://terapix.iap.fr/rubrique.php?id\_rubrique=221} and STUFF
1.17\footnote{http://terapix.iap.fr/rubrique.php?id\_rubrique=248} developed by
\citet{skymaker}. The galaxy distribution was created using STUFF
1.17 default parameters for a 1024$\times$1024 image and represent the galaxy
sample of interest for which we wish to measure the photometry. The instrument
specific parameters (e.g. pixel size, mirror size) were set to GALEX
specifications. The galaxy distribution includes objects with
18$\le$MAG\_LIMITS$\le$26.5. We then created a simulated image with SkyMaker
for a $\sim$33 ksec exposure using the GALEX NUV PSF. The comparison in
number density and brightness distribution between the real and the simulated
GALEX NUV data shows good agreement (Fig.~\ref{simuhist}). We added a
uniformly distributed sample of 100,000 artificial background 
galaxies to the data, simulated using {\it IRAF} tasks {\it gallist} and {\it
  mkobjects}. The sample was made using a power law
distribution $dN \sim m^{-\beta}$ with $\beta = 0.6$, 
restricted to 18~$\le m\le$~28.5 including background galaxies down to the 1\%
flux level of out NUV magnitude limit. Comparison of SExtractor catalogs with
and without the synthetic galaxies shows no significant effects on the
photometric results for objects with $m\le23.5$ (see Fig.~\ref{simuconf}). The
rms for the change in m$_{iso}$ is dm$_{rms}$ = 0.0707~mag,
including objects which have $\rm dm < 0$ resulting from changes in the
deblending during the SExtractor search and/or overall higher background level
due to the high number of background galaxies. The rms of change in m$_{iso}$
accounting only for objects with $\rm dm \ge 0$ is dm$_{rms}$ = 0.0784. This
is small compared to the NUV scatter in the averaged SEDs, which can be as
high as 1.5 mag.   
We conclude that our catalogs are 80 - 90,\% 
complete down to m$_{NUV}$ = 24\,AB. For LBG candidates we are 40 - 60\,\%
complete to AB$_{NUV}$ = 24.5.  

\section{Photometric redshifts}
\label{redshifts}

\begin{figure*}
\epsscale{1.1}
\plottwo{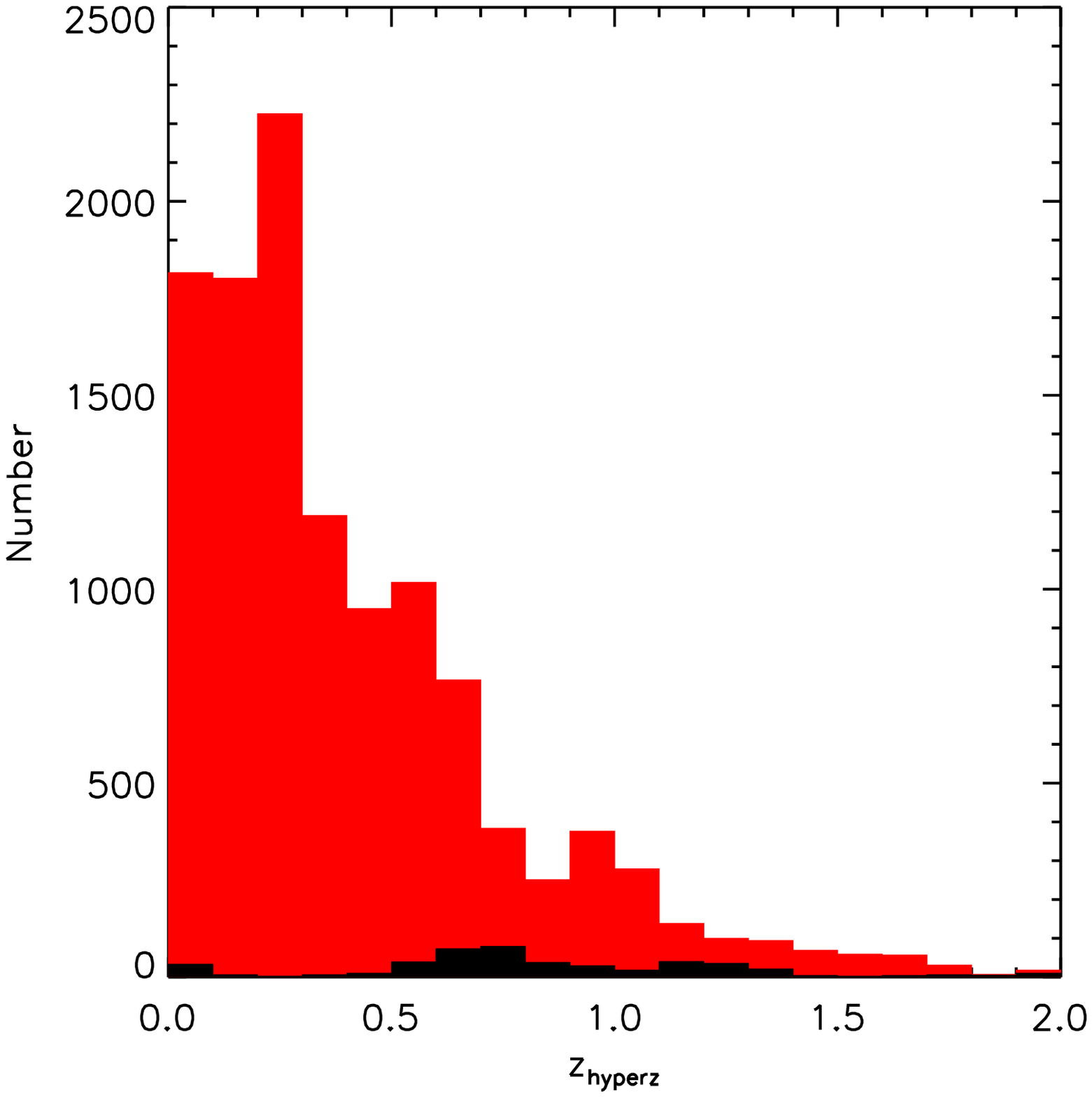}{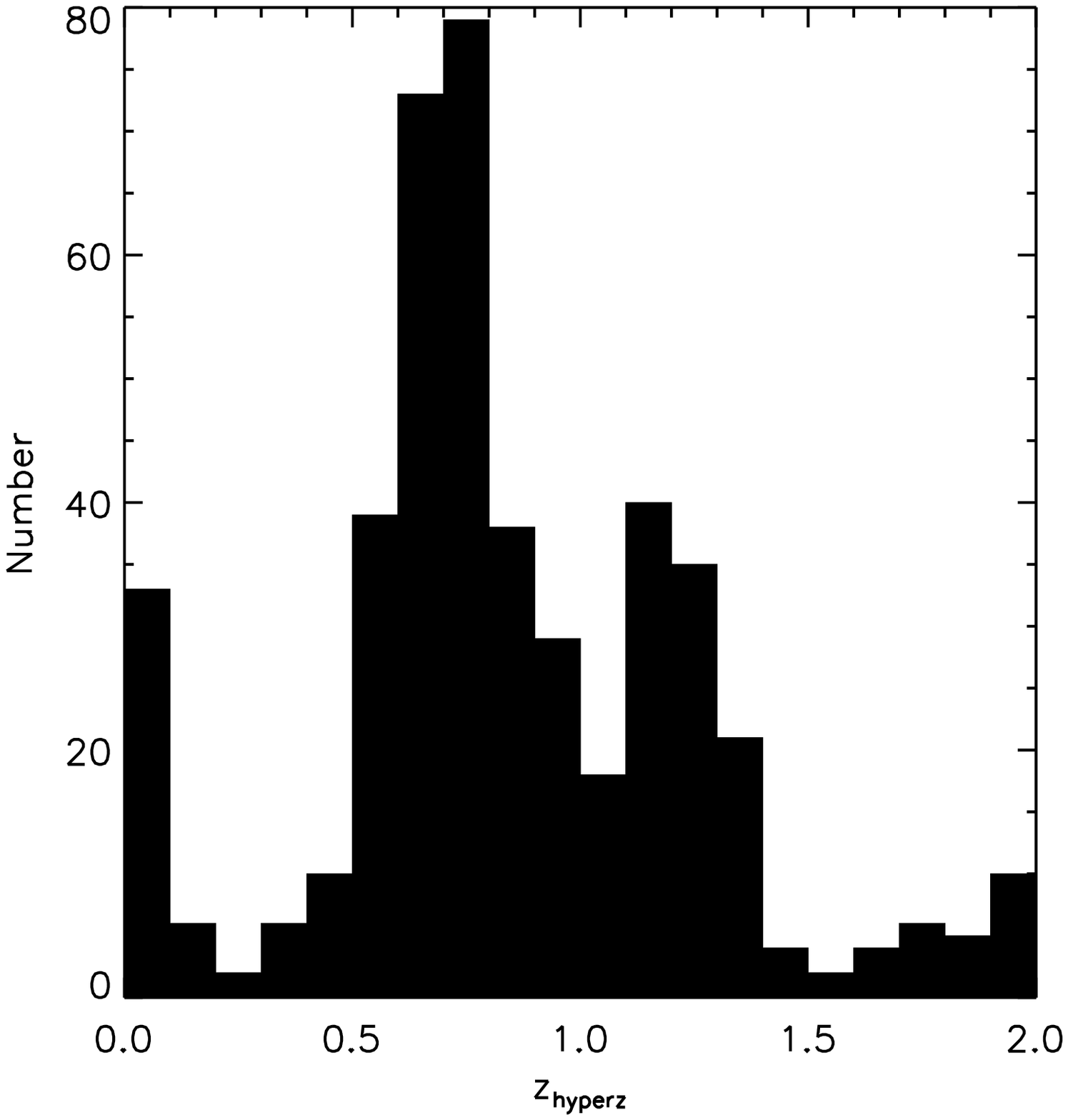}
\caption{Photometric redshift distribution for all galaxies (left, red) and
  the LBG candidates (left and right black). We used a bin size for the
  redshift of $\delta$z=0.1. The red bars (left panel) represent the
  distribution in redshift for all galaxies for which we also have SDSS
  counterparts, while the black filled bars (left+right panel) display the
  photometric redshift distribution for the LBG candidates.}
\label{redhisto}
\end{figure*}

\placefigure{redhisto}
We used seven band photometry from the two GALEX (FUV+NUV) and
the five SDSS DR5 (u,g,r,i,z) filter bands for photometric redshifts of our
final UV selected sample, applying the algorithm from {\it hyperz v1.1}
\citep[][]{2000A&A...363..476B}. The variance of Lyman-$\alpha$ opacity in the
intergalactic medium \citep[][]{2001A&A...380..425M} produces a negligible
effect at these redshifts. The redshift determination is done by
cross-correlating a set of template spectra to the colors of the sample
galaxies. In the current version of {\it hyperz} we used
a set of four template spectra from \citet[][]{2003MNRAS.344.1000B},
consisting of an elliptical, Sc, Sd galaxy, and star-burst. The resulting
redshift distributions for the total galaxy and LBG candidate samples are
shown in Fig.~\ref{redhisto}.  

\begin{figure}
\epsscale{1}
\plotone{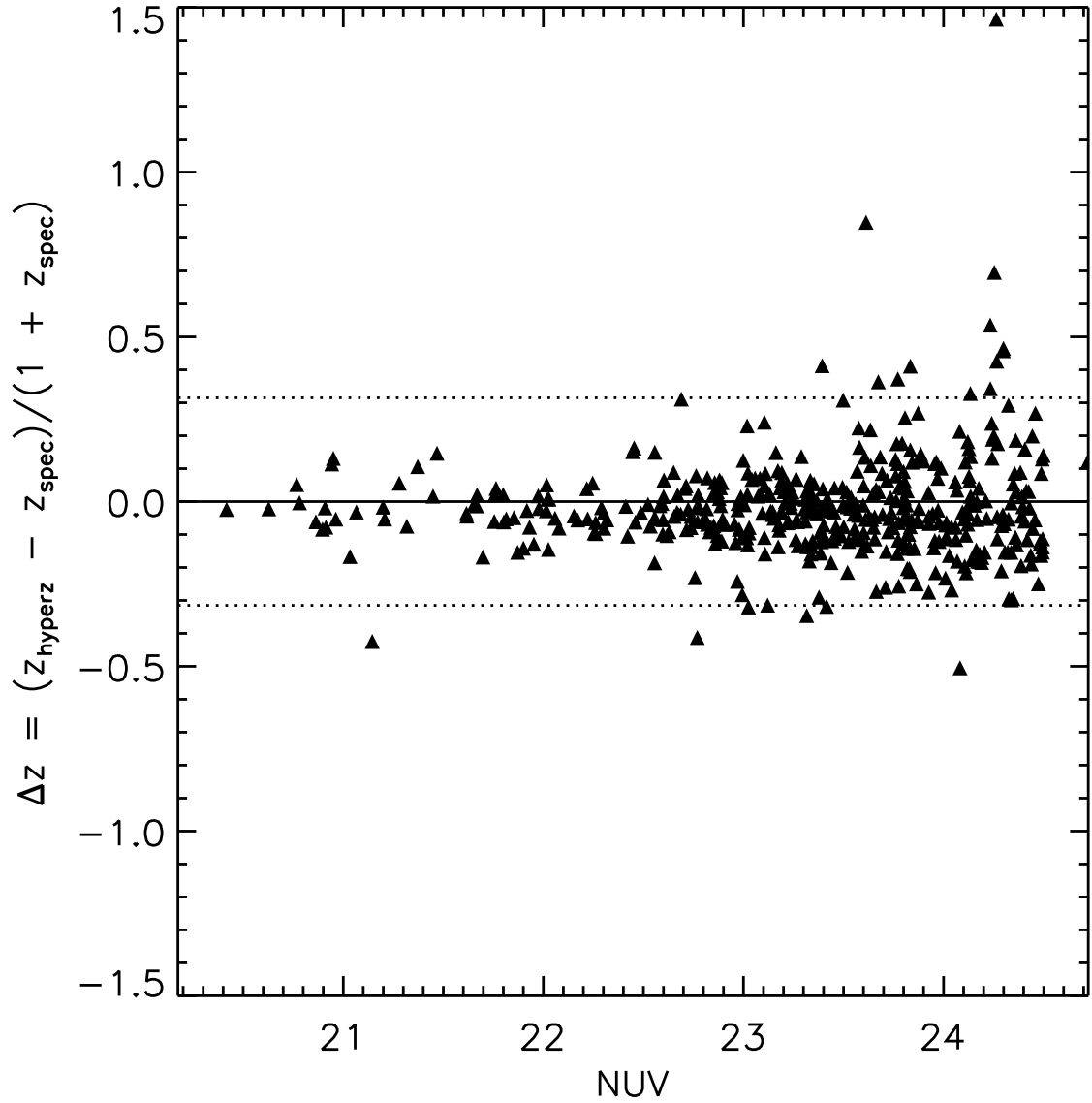}
\caption{Total NUV magnitude in relation to the
  photometric redshift accuracy $\Delta z$. The dashed lines represent the
  3$\sigma$ deviation from the $\Delta z$\,=\,0 line. For galaxies fainter than
  23.5\,mag, the photometric redshift accuracy decreases significantly.
}
\label{photoz2}
\end{figure}

\begin{figure*}
\epsscale{1}
\plottwo{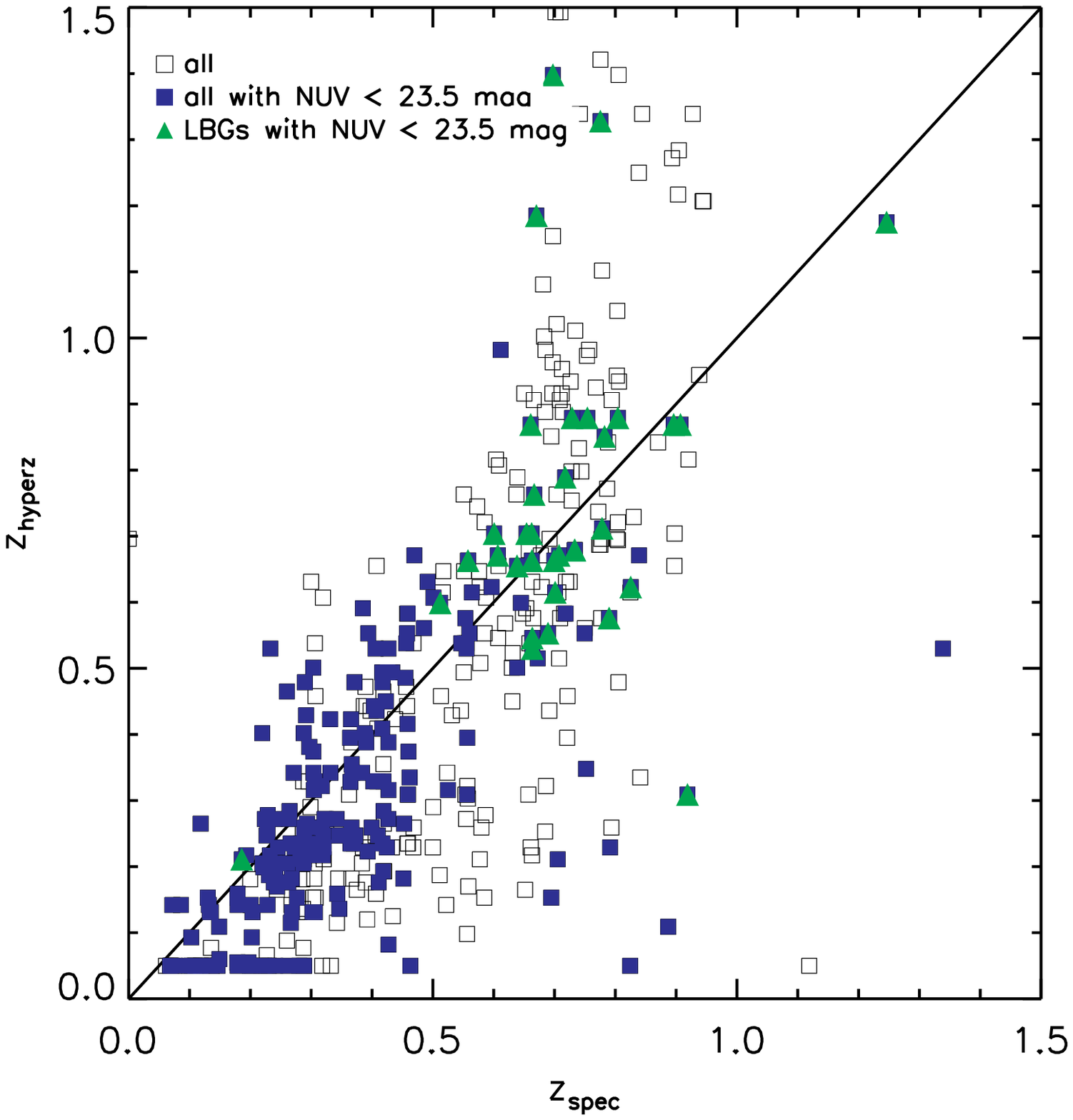}{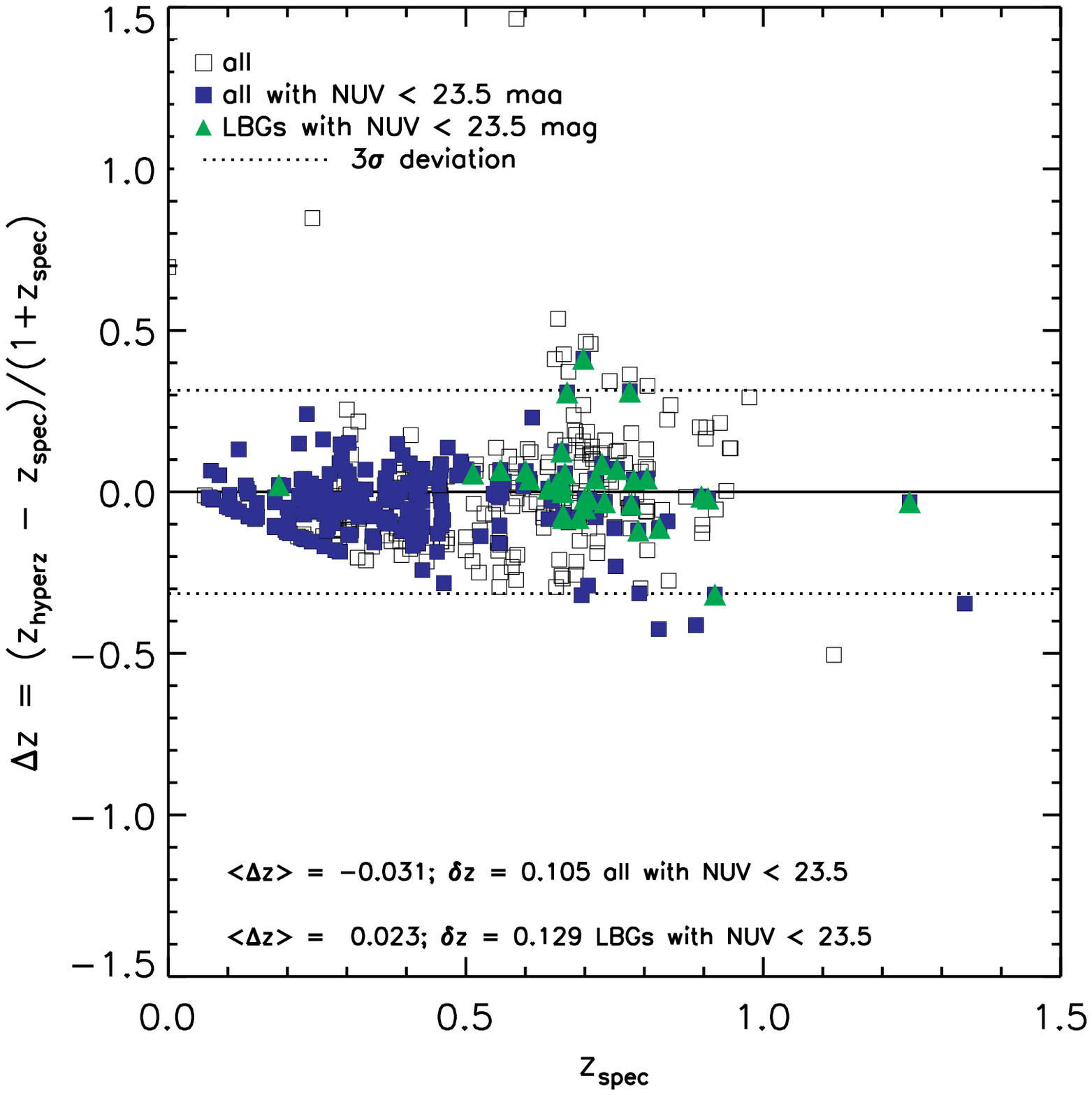}
\caption{The left panel shows the comparison of our photometric redshifts
  $z_{hyperz}$ obtained using 7 band photometry (FUV,NUV,u,g,r,i,z) to
  spectroscopic redshifts which we obtained for a subsample of 448
  galaxies. Blue triangles are galaxies with m$_{NUV}\,\le\,$23.5\,mag. Green
  triangles represent the LBG candidates with m$_{NUV}\,\le\,$23.5\,mag. The
  right panel shows the residuals $\Delta$z using the same color schemes. From
  the right panel we see that the number of catastrophic outliers is
  relatively small (all 2.6\% and LBG candidates 5.9\%) for galaxies with
  m$_{NUV}\,\le\,$23.5.  
}
\label{photoz}
\end{figure*}

\placefigure{photoz2}
To get information about the accuracies of our photometric redshifts
estimation, we compared the resulting photo-z's of the final galaxy catalog
with a subsample of 448 galaxies for which we have spectroscopic redshifts
(see Figs.~\ref{photoz2} and \ref{photoz}).  
The spectra were observed using the IMACS mult-object
spectrograph at the 6.5~m Baade Magellan telescope \citep[][]{harris09}.
The spectroscopic subsample covers 0.06$<$z$<$1.34 with a mean redshift of
$\left < z \right >$ = 0.48$\pm$0.23. 
The photometric redshift accuracy decreases
significantly for objects with m$_{NUV} >$23.5~mag  (Fig.~\ref{photoz2}),
and the standard
deviation $\sigma_{\Delta z}$ increases from 0.105 (0.129 for LBG candidates)
to 0.203 (0.195 for LBG candidates). We therefore restricted our analyses to
objects brighter than m$_{NUV}$\,=\,23.5 (blue and green triangles in
Figs.~\ref{photoz2} and \ref{photoz}). This leaves a more conservatively 
selected sample of the 462 LBG candidates.   
For our spectroscopic subsample we derived mean offsets of $\left<
  \Delta z \right> = -0.031$ for the whole  bright sample (blue + green
triangles) and $\left< \Delta z\right> = 0.023$ for the LBG candidate sample
(green triangles). We therefore see no significant systematic offsets.
The fraction of catastrophic outliers ($\Delta z\,>\,3\sigma$,
Fig.~\ref{photoz}) for bright
objects (m$_{NUV}\le$23.5) is about 2.6\% (5.9\% for LBG candidates). 
\placefigure{photoz}

Using our GALEX plus 7 band SDSS photometry, we are able to obtain
unbiased photometric redshifts with well defined uncertainties for objects
with NUV$\le$23.5 for both the complete galaxy as well as the LBG
sample. Hence our LBG selection should be reasonably robust, given the depth in
redshift of our four galaxy samples (see \S~\ref{sedfit} and \ref{lbgsamp})
which is 2-3\,$\sigma$ for the derived uncertainties of our photometric
redshifts. We therefore assume no significant impact from the relatively
large photo-z scatter on our results.

A listing of our LBG sample is in Table~\ref{lbg_example}, available in the
electronic edition of this paper.  We also provide information on the LBGs on
our team website\footnote{{\tt http://www.physics.uofl.edu/lqg}}.
\placetable{lbg_example}

\section{SED fitting}
\label{sedfit}

To constrain the evolution of galaxies in the dense quasar environment of the
two LQGs, we compared averaged SEDs built out of the photometric measurements of
our sample galaxies to model SEDs derived from the synthesis evolution model
PEGASE \citep[][]{1997A&A...326..950F}. We used a library
of several thousand PEGASE spectra from an extended parameter
study to probe the star formation histories (SFHs) of a sample of Low Surface
Brightness (LSB) galaxies \citep[][]{haberzettl07}. 

We made stacked SEDs of our LBG candidate samples (\S~\ref{lbgsamp})
from the GALEX+\linebreak SDSS 7-band photometry. The stacked SEDs were
constructed for 
four different redshift bins over $0.5\le z<0.7$
(foreground sample FG1), $0.7\le z<0.9$ (LQG0.8), $0.9\le z<1.2$ (foreground
sample FG2) and $1.2\le z<1.5$ (CCLQG). To reduce the influence of
extreme objects (e.g. redshift outliers, mis-identified stars) we applied
upper flux limits for the averaged SEDs in all filter bands, excluding
iteratively all objects which have fluxes more than 3$\sigma$ from the mean
flux.  
Additionally, we only used objects which have non-negative fluxes in all filter
bands. The errors for the mean fluxes in the single filter-bands are
represented by their standard deviations. 
By fitting the model SEDs to averaged low resolution spectra of our LBG
sample, we were then able to obtain SFHs and luminosity-weighted ages
(results described in \S~\ref{lbgsamp}). The model SED
library contains spectra calculated for different star formation laws
(SFLaws), star formation rates (SFRs) and extinction
geometries. The model SEDs were constructed accounting for consistent chemical
evolution. This is a more realistic approach than the use of simple stellar
populations (SSPs) with fixed metallicities, giving us the advantage that the
metallicity is not a free parameter. Therefore, we preferred PEGASE
over other models as for example the widely used
\citeauthor{2003MNRAS.344.1000B} models \citep[][]{2003MNRAS.344.1000B}. 
We chose four different SFLaws including star-burst
scenarios, constant, exponentially decreasing, and power law SFRs.  
The best matching SEDs were derived by performing a $\chi^2$-fit between
model and measured SEDs. 
To judge the quality of our SED fit results we compare the reduced
$\chi^2$-values against the ages of the fitted SEDs (Figs.~\ref{chi2_age},
\ref{chi2_age_ext}, \ref{chi2_age_blue_red}, \ref{chi2_age_blue_red_ext}),
which is the 
parameter of main interest to us. From these plots we estimate the 1$\sigma$
uncertainties by comparing $\chi^2$ for every SED to $\chi^2_{min} + 1$.  

Since we are analyzing samples of LBG candidates, we set the measured
fluxes in the GALEX FUV band to zero for those redshift bins where
the FUV-band would be below 912\,\AA\, in the rest frame.
We normalized the flux of both modeled and measured
SEDs with respect to the rest frame flux in z-band, and
redshift-corrected the measured SEDs according to the mean redshift of the
group to which each of the averaged LBG spectra belongs. We next describe
  the stacked SEDs and results.

\section{Lyman Break Galaxy candidate sample}
\label{lbgsamp}

 Our NUV$<$23.5 criterion gives a conservative sample of 462 LBGs, for
 which our 7-band UV-optical photometric redshift distribution is shown in the
 two histograms in Fig.~\ref{redhisto}. The red filled bars represent the
 photometric redshift distribution for the whole galaxy sample, while the LBG
 candidate sample is represented by the black filled bars. From the right
 panel we see that most (405) of the LBG candidates have photometric
 redshifts $z\ge$0.5. The mean photometric redshift of the LBG
 candidate sample is $\left<z_{LBG}\right>=0.86\pm 0.45$, compared to the mean
 of the final galaxy sample of $\left<z_{galaxies}\right>=0.41\pm 0.35$. Only
 60 LBG candidates lie between 1.2$\le$z$<$1.5 and are associated
 with the CCLQG. We mainly probe the foreground LQG (LQG0.8) at
 0.7$\le$z$\le$0.9 (117 LBG candidates). The FUV dropout technique effectively
 identified $z\ge$0.5 galaxies: 88~\% are at $z\ge$0.5.  

\begin{figure*}
\epsscale{1}
\plotone{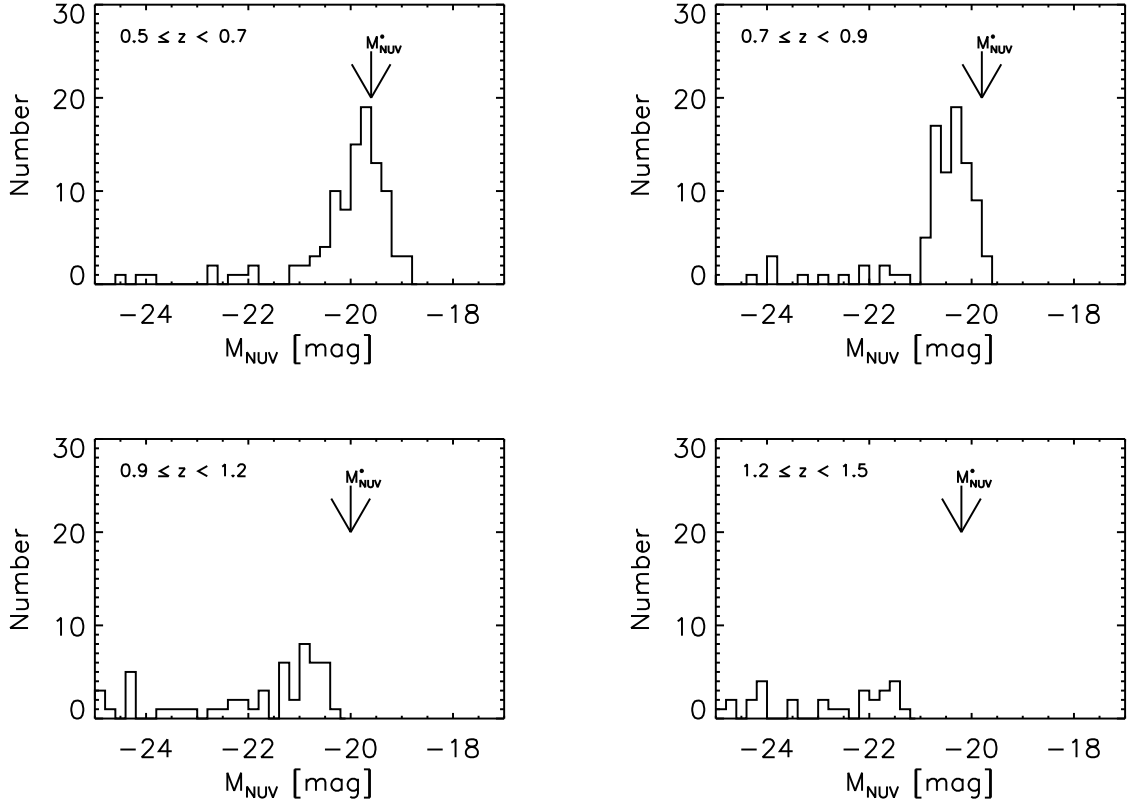}
\caption{Distribution of absolute NUV magnitudes (rest frame FUV) used to
  select the bright and faint LBG subsample. The selection limits are
  indicated by the values for the M$^*_{NUV}$ of 
  \citet[][M$^*_{NUV}$ = -19.6, -19.8, -20.0, and -20.2 from low to high
 redshift]{2005ApJ...619L..43A} at the individual redshifts.}
\label{nuvhisto}
\end{figure*}

\placefigure{nuvhisto} 
For further analysis, we have divided the LBG candidate sample into four
 redshift bins (Table~\ref{subsample}): 0.5-0.7 (FG1 sample), 0.7-0.9
 (LQG0.8 sample), 0.9-1.2 (FG2 sample), and 1.2-1.5 (CCLQG). For each redshift
 bin, we selected two subsamples for intrinsically bright and faint LBG
 candidates according to their absolute brightness in the GALEX NUV filter
 band (rest frame FUV). For
 the selection we used the values of M$^*_{NUV}$ from
 \citet[][M$^*_{NUV}$ = -19.6, -19.8, -20.0, and -20.2 from low to high
 redshift]{2005ApJ...619L..43A} consistent with the four redshift
 bins, and K-corrected the measured NUV
 magnitudes using the {\it kcorrect} software v4\_1\_4 of \citet[][see also
 Fig.~\ref{nuvhisto}]{2003AJ....125.2348B}. The bright and faint subsamples
 consist of 3 to 69 candidates. For the FG2 and CCLQG sample, we were not able
 to detect any faint (M$^*_{NUV} \ge$-19.8~mag or -19.6~mag) LBG candidates. 

 Finally, we divided our LBG candidate sample into red and blue
 subsamples using the MSL in the g$-$i vs. r$-$i color-color
 diagram. For all redshift bins, this selection resulted in a larger red than
 blue subsample and is a first indication that our LBG samples are dominated
 by either dusty or more evolved galaxies. 
\placetable{subsample}
 
\subsection{Star Formation History}

\begin{figure*}
\figurenum{10a}
\epsscale{0.8}
\plotone{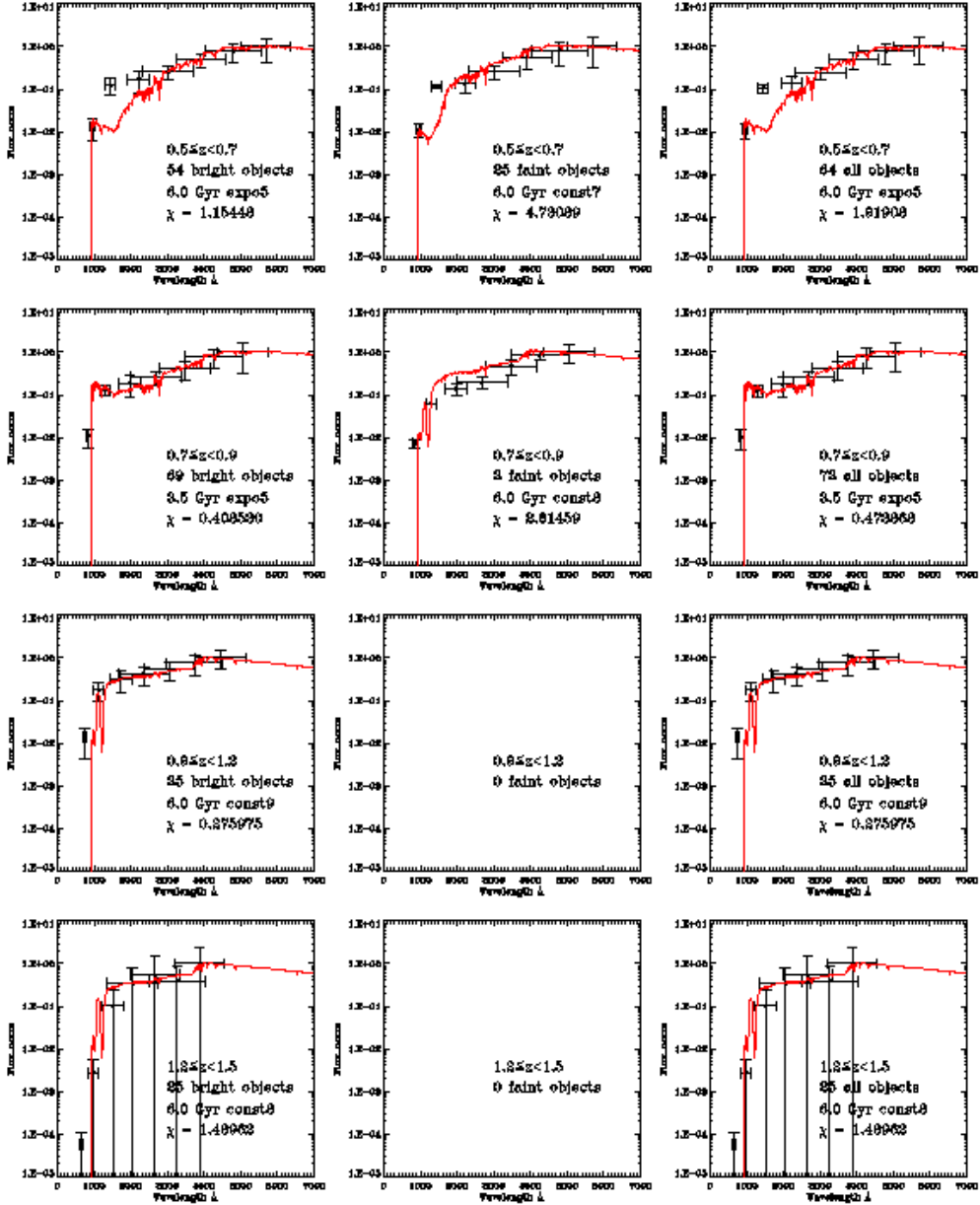}
\caption{\label{lbg_sfh}Averaged spectral energy distribution of the LBG
  candidates (black) found in the GALEX data. The LBG candidates are fitted
  with PEGASE models without dust. The samples are divided into bright
  (left column) and faint (middle column) subsample. The right column shows
  the results for the combined sample.  We also divided the LBG sample into 4
  redshift bins from $z\ge$\,0.5 (top) to $z<$\,1.5 (bottom). The
  green/light lines represent the best fitting model SEDs.  
}
\end{figure*}

\begin{figure*}
\figurenum{10b}
\epsscale{0.8}
\plotone{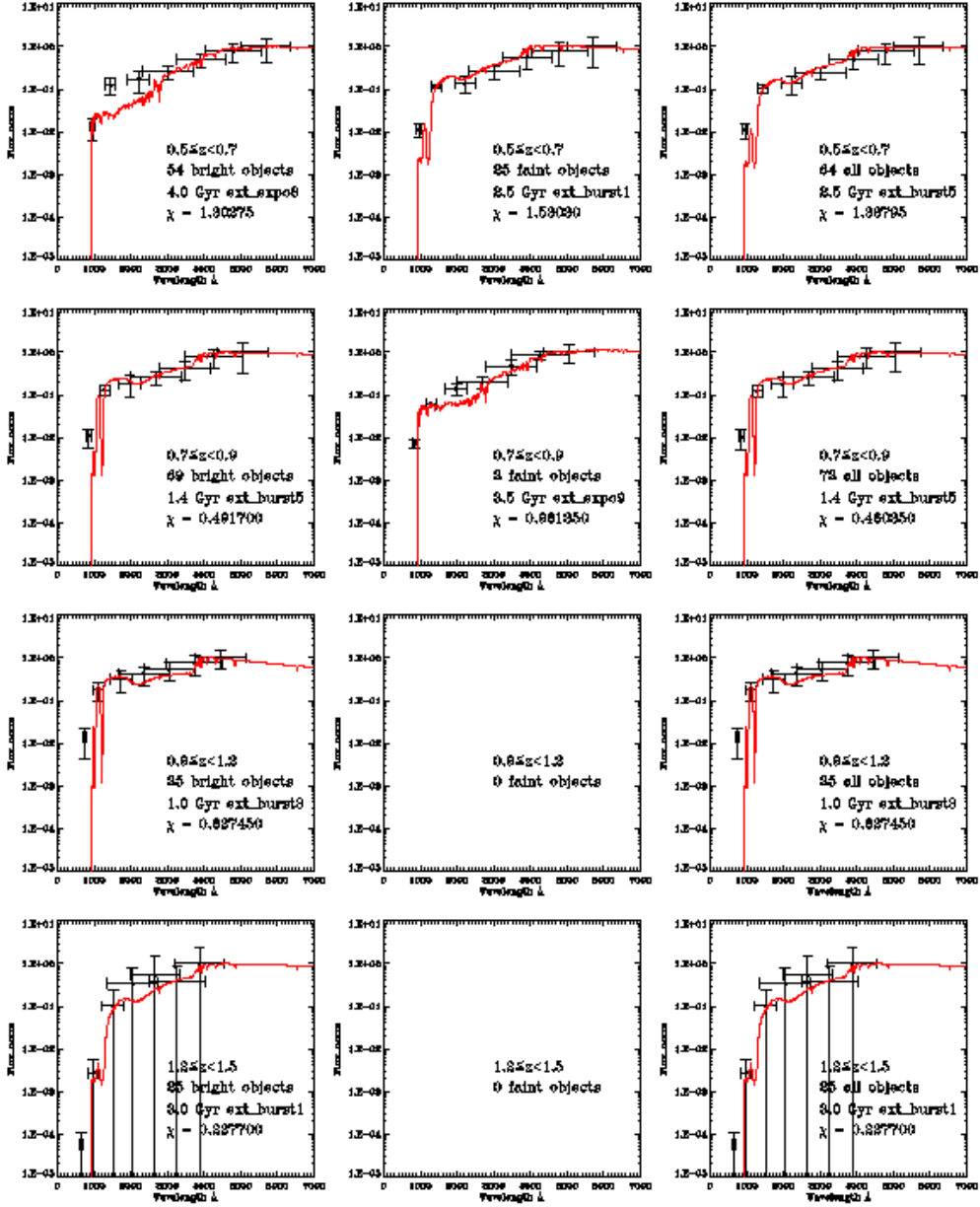}
\caption{\label{lbg_sfh_ext} The same as Fig.~\ref{lbg_sfh}, except the
  inclusion of dust in the PEGASE models. Samples are again divided into
  bright (left column) and faint (middle column) subsamples. The right column
  shows the results for the combined samples. From top to bottom we have
  again plotted the results for the different redshift bins. 
}
\end{figure*}

\begin{figure*}
\figurenum{11a}
\epsscale{0.8}
\plotone{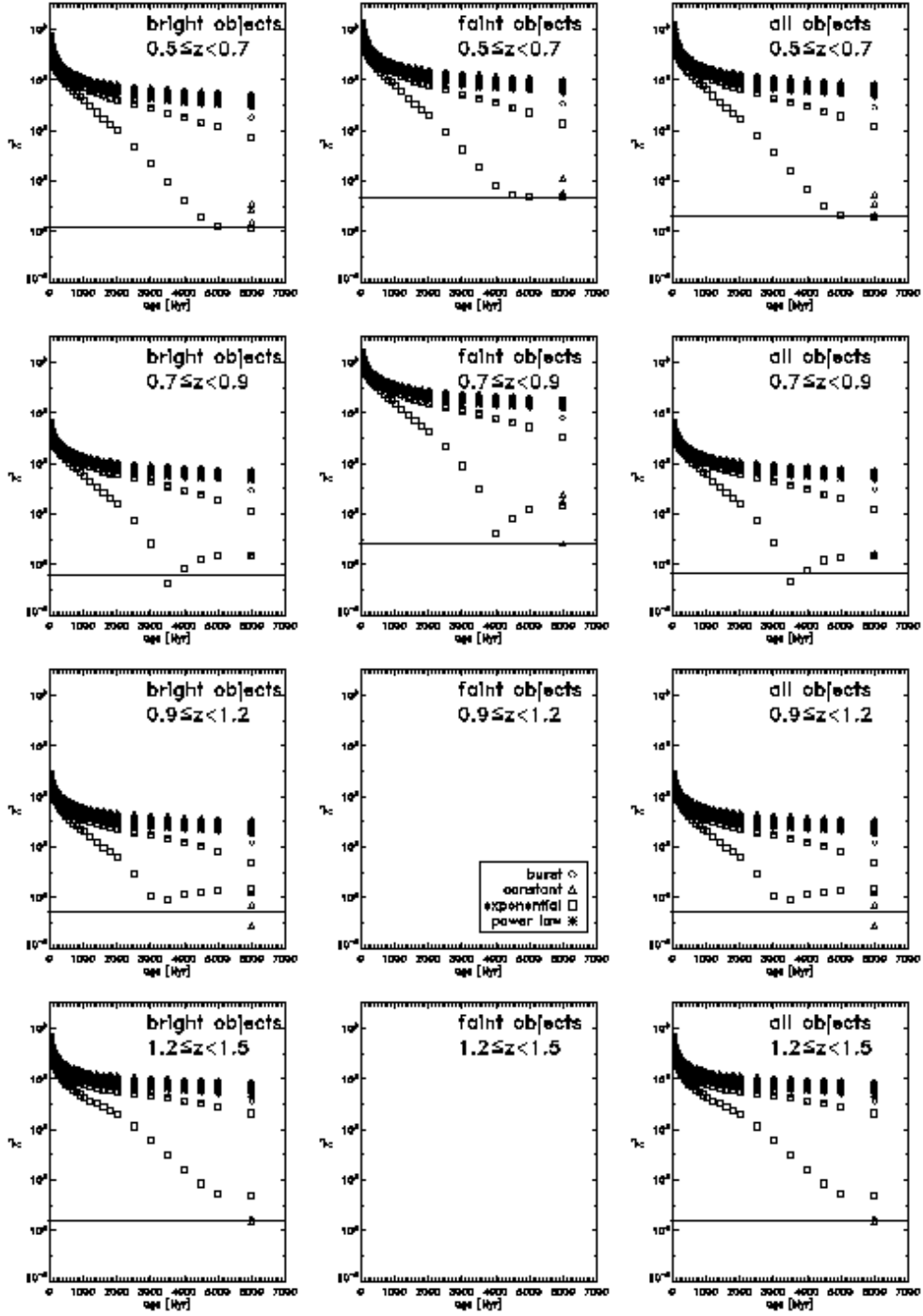}
\caption{\label{chi2_age} Results of the model fitting (see
  Fig.~\ref{lbg_sfh}) indicated by the
  reduced $\chi^2$ vs. age in Myr for bright and faint subsamples as well as the
  complete galaxy sample according to their redshift intervals. The
  different symbols indicate fit results for different star formation models
  used (burst: diamonds; constant: triangles; exponential: squares; power law:
  asterisks). The horizontal lines represent the 1$\sigma$ limits in $\chi^2$
  for the SED fits.}
\end{figure*}

\begin{figure*}
\figurenum{11b}
\epsscale{0.8}
\plotone{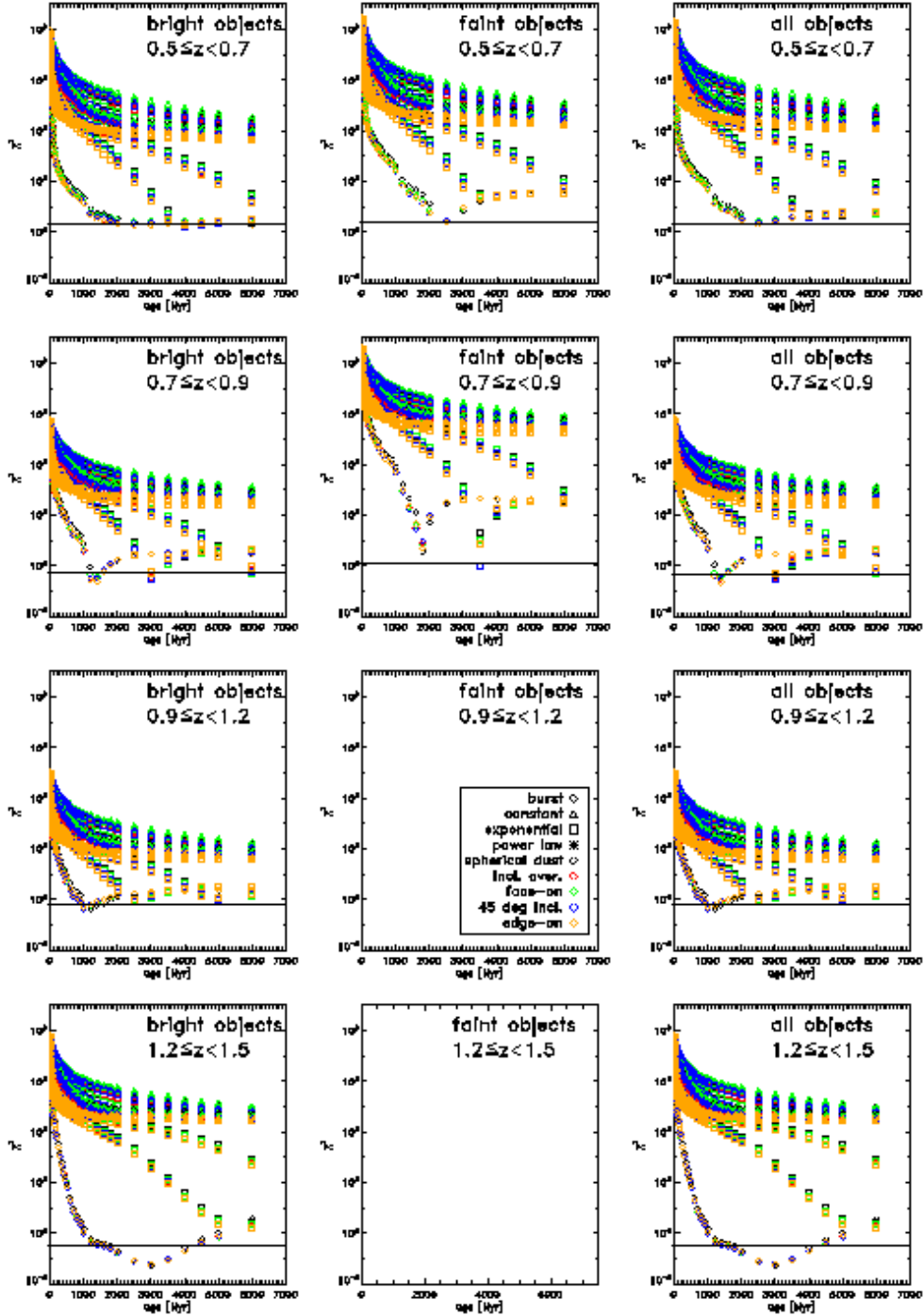}
\caption{\label{chi2_age_ext}The same as Fig.~\ref{chi2_age}, except the
  inclusion of dust in the PEGASE models. Again different symbols represent
  different star formation models (burst: diamonds; constant:
  triangles; exponential: squares; power law: asterisks),
  while the different colors 
  indicate the use of different dust geometries (spherical: black; disk
  geometry: inclination averaged -- red; face-on -- green; 45$^{\circ}$ --
  blue; edge-on -- orange). The solid lines again represent the $1\sigma$
  limits for SED fits (see Fig.~\ref{lbg_sfh_ext}).}
\end{figure*}

\placefigure{lbg_sfh}
\placefigure{chi2_age}
\placefigure{lbg_sfh_ext}
\placefigure{chi2_age_ext}
We used the SED fits of \S~\ref{sedfit}
(Figs.~\ref{lbg_sfh},b) to constrain the LBG candidate star
formation histories. Although $\chi^2$-fits did not result in unique
solutions (see
Figs.~\ref{chi2_age},b)
the best fitting SEDs give luminosity-weighted 
ages between 3.5 - 6 Gyr for the dust-free
models and 1.0 - 4.0 Gyr for models including dust.
Results for the SED fitting are summarized in
Tables~\ref{sedfitres_noext}-\ref{sedfitresext_color} and
explained in more detail below.
\placetable{sedfitres_noext}
\placetable{sedfitres_ext}

 The best fitting SFH for the FG1 sample, consisting of 64 LBG candidates
 with \linebreak-20.38$\le$M$_{NUV}$$\le$-18.85, is described by an
 exponentially decreasing SFR (decay time $\tau = 0.5$~Gyr) after 6~Gyr for
 the dust-free and a star-burst after 2.5~Gyr for dust-containing models using
 an edge-on disk geometry. Subdividing FG1
 into a bright (M$_{NUV} <$ -19.6) and faint (M$_{NUV} \ge$ -19.6) subsample
 resulted in the same SFHs for the dust-free model. For the models including
 dust the bright subsample is best represented by a 4.0~Gyr old exponentially
 decreasing SFR (decay time $\tau = 1.0$~Gyr) assuming a face-on disk geometry
 for the dust distribution. The faint subsample is best described by a
 star-burst scenario after 2.5~Gyr using an edge-on disk geometry. The
 dust-free and dusty models for the complete sample and the dust-free models
 for the bright and faint subsamples are both relatively well constrained,
 allowing for only one solution within the $\chi^2 + 1$ limit. 
 The ages derived from the dusty models for the bright subsample range
 between 2.5 and 6 Gyr for a star-burst and exponentially decreasing
 SFRs with different dust geometries. The faint subsample is fitted with
 2.5~Gyr old star-bursts with spherical, face-on and edge-on dust
 geometries.  

 The fits for the FG2 sample ($ 0.9\le {\it z} <1.2$) resulted in SFHs with
 slightly younger 
 luminosity-weighted ages. The total sample included in the fits consists of
 35 LBG candidates with -22.28\,$\le$\,M$_{NUV}$\,$\le$\,-20.38 and could be
 best fitted by a constant SFR after 6.0~Gyr for the dust-free (only fit
 acceptable) and a burst
 scenario after 1.0~Gyr for the dusty model. The dust is
 assumed to be distributed with a face-on disk geometry. For this redshift
 slice we were not able to detect a faint subsample. The acceptable SEDs
 including dust result in star-bursts with ages ranging between 1 to 1.4~Gyr
 using the five different dust geometries in the library (spherical,
 inclination averaged, face-on, 45$^{\circ}$ inclined, and edge-on). 

 The foreground large quasar group (LQG0.8 sample) consists of 73 LBG
 candidates with absolute magnitudes -20.97$\le$M$_{NUV}\le$-19.67. The SFH is
 best described by an exponentially decreasing SFR (decay time $\tau =
 0.5$~Gyr) after 3.5~Gyr for the dust-free and a star-burst after 1.4~Gyr for
 the dusty models (edge-on disk). For
 the bright (M$_{NUV} <$ -19.8) and faint (M$_{NUV} \ge$ -19.8) subsamples the
 SFHs were best fitted by exponentially decreasing SFRs (decay time $\tau =
 0.5$~Gyr) after 3.5~Gyr (bright) and constant SFR after 6~Gyr (faint)
 assuming dust-free models. The models including dust extinction were best
 fitted by a star-burst after 1.4~Gyr (bright) and an exponentially
 decreasing SFR (faint, decay time $\tau = 0.5$~Gyr) after 3.5~Gyr
 respectively. The dust distributions are assumed to represent edge-on
 (bright) and 45$^\circ$ inclined disk (faint) geometries.  
 The dust-free models for the complete, bright, and faint subsamples are
 well constrained, allowing for one fit within the $1\sigma$ limit. For
 the models including dust, the acceptable fits result in ages between
 1.2 and 3 Gyr for the complete and 1.2 and 6 Gyr for the faint susbample
 using a star-burst for the complete and an exponentially decreasing SFR for the
 faint subsample including all dust geometries. The bright subsample is well
 represented by exponentially decreasing SFR after 3.5 Gyr using an inclination
 averaged or 45$^{\circ}$ disk dust geometry.  

\placetable{sedfitres_color}
\placetable{sedfitresext_color}
The CCLQG LBG sample only consists of 25 LBG candidates with absolute
magnitudes M$_{NUV} \ge$ -21.34. Therefore, we could only derive SFHs for the
M$>$M$^*$ LBG candidates. For dust-free models, the SFH is best
described by a constant SFR after 6~Gyr (only acceptable solution), while for
the models including extinction the best fit is from a star-burst after 3~Gyr
(spherical dust geometry).
The dusty models also allow for solutions ranging from 1.6 to 4 Gyr for
the age, using star-bursts with all five possible dust geometries.

The SFHs for all subsamples resulted in significantly older best fitting ages
compared to the results of \citet[][]{2007MNRAS.380..986B}. The ages
derived here correspond to formation redshifts between 1.5 and 5, which
is consistent with the peak of the luminosity density in the
Universe \citep[e.g.,][]{2000ApJ...541...25N,2006ApJ...648..299S}.  
Although we have no direct measurement of the
dust content for our LBG candidates, the dusty models fit best the LBG
candidate samples in the two LQGs (LQG0.8 and CCLQG) indicating that dust
plays a non-negligible role.

\begin{figure*}
\figurenum{10c}
\epsscale{0.8}
\plotone{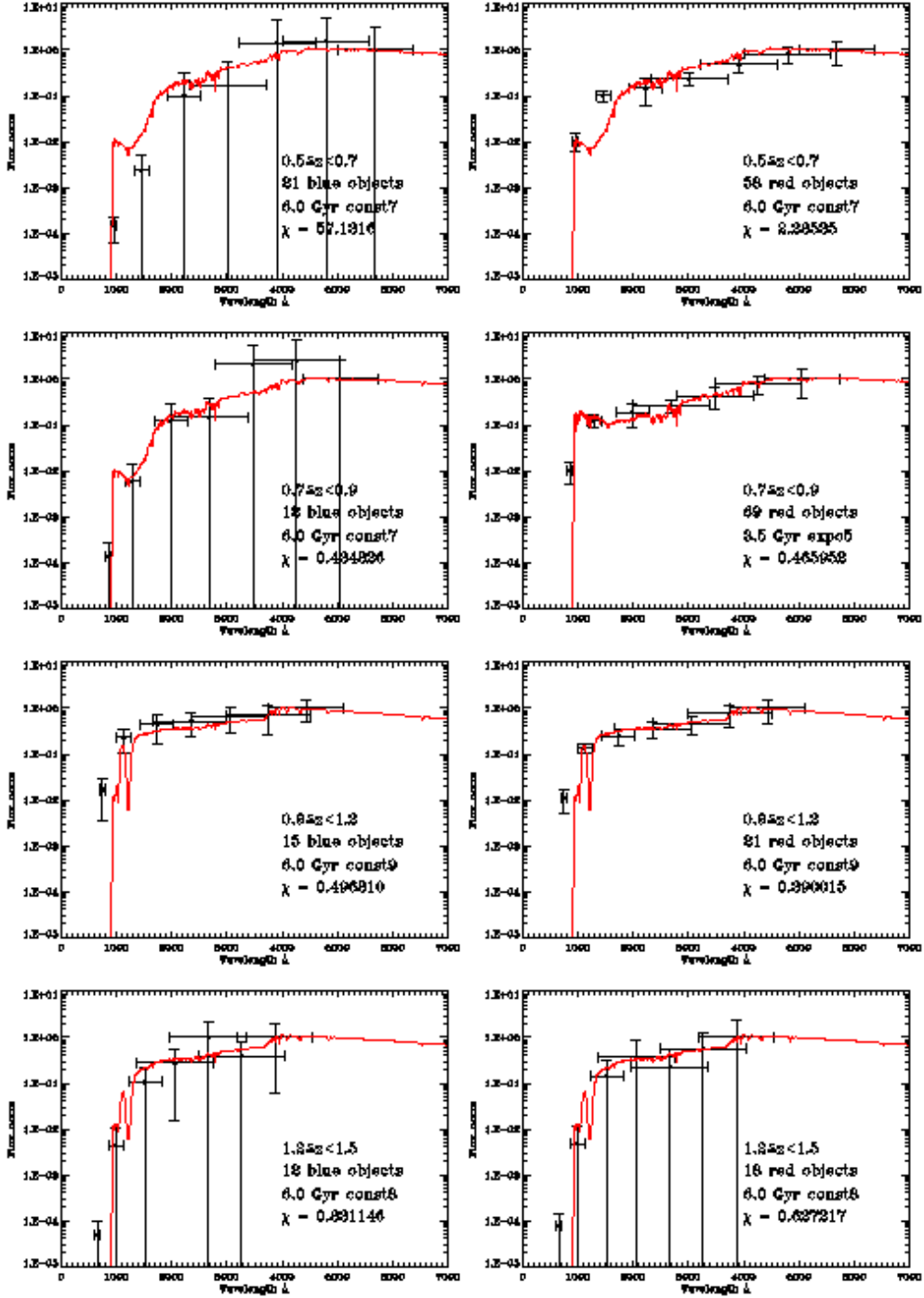}
\caption{\label{lbg_blue_red_sfh} The diagram shows the result for the SED
  fitting for blue (left column) and red (right column) subsamples selected
  using the MSL as defined by \citet[][see
  Fig.~\ref{starsel}]{2007AJ....134.2398C}. The green lines represent
  the best fitting dust-free model SEDs. We again divided the sample
  into four subsamples according to the redshift bins from low (top) to high
  (bottom) redshift.}
\end{figure*}

\begin{figure*}
\figurenum{10d}
\epsscale{0.8}
\plotone{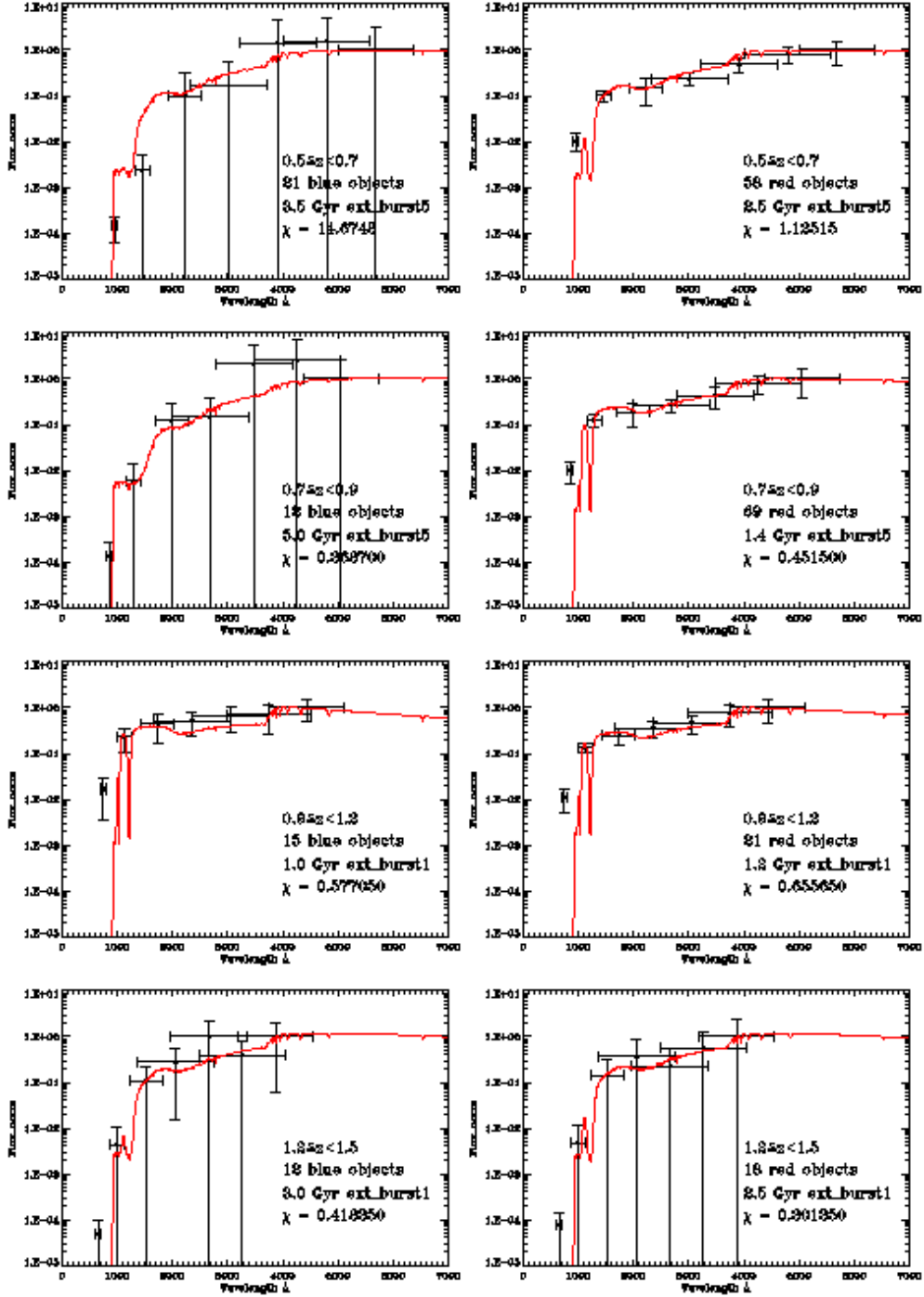}
\caption{\label{lbg_blue_red_sfh_ext} The same as Fig.~\ref{lbg_blue_red_sfh},
  except the inclusion of dust in the PEGASE models.  
}
\end{figure*}

\begin{figure*}
\figurenum{11c}
\epsscale{0.8}
\plotone{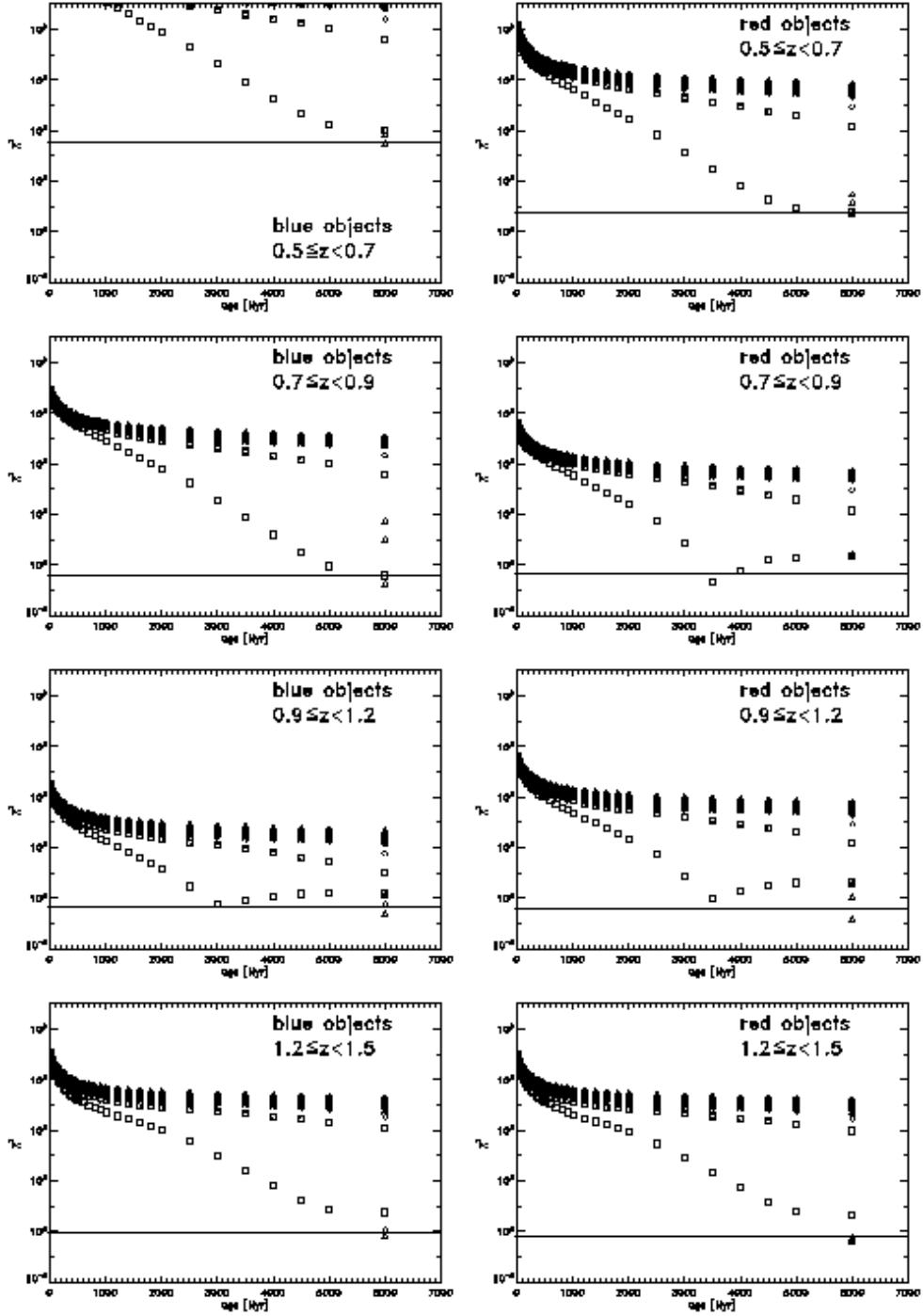}
\caption{\label{chi2_age_blue_red}Results for the SED model fits of the blue
  and red subsamples (see Fig.~\ref{lbg_blue_red_sfh}). The use of symbols is
  the same as in Fig.~\ref{chi2_age}.}
\end{figure*}

\begin{figure*}
\figurenum{11d}
\epsscale{0.8}
\plotone{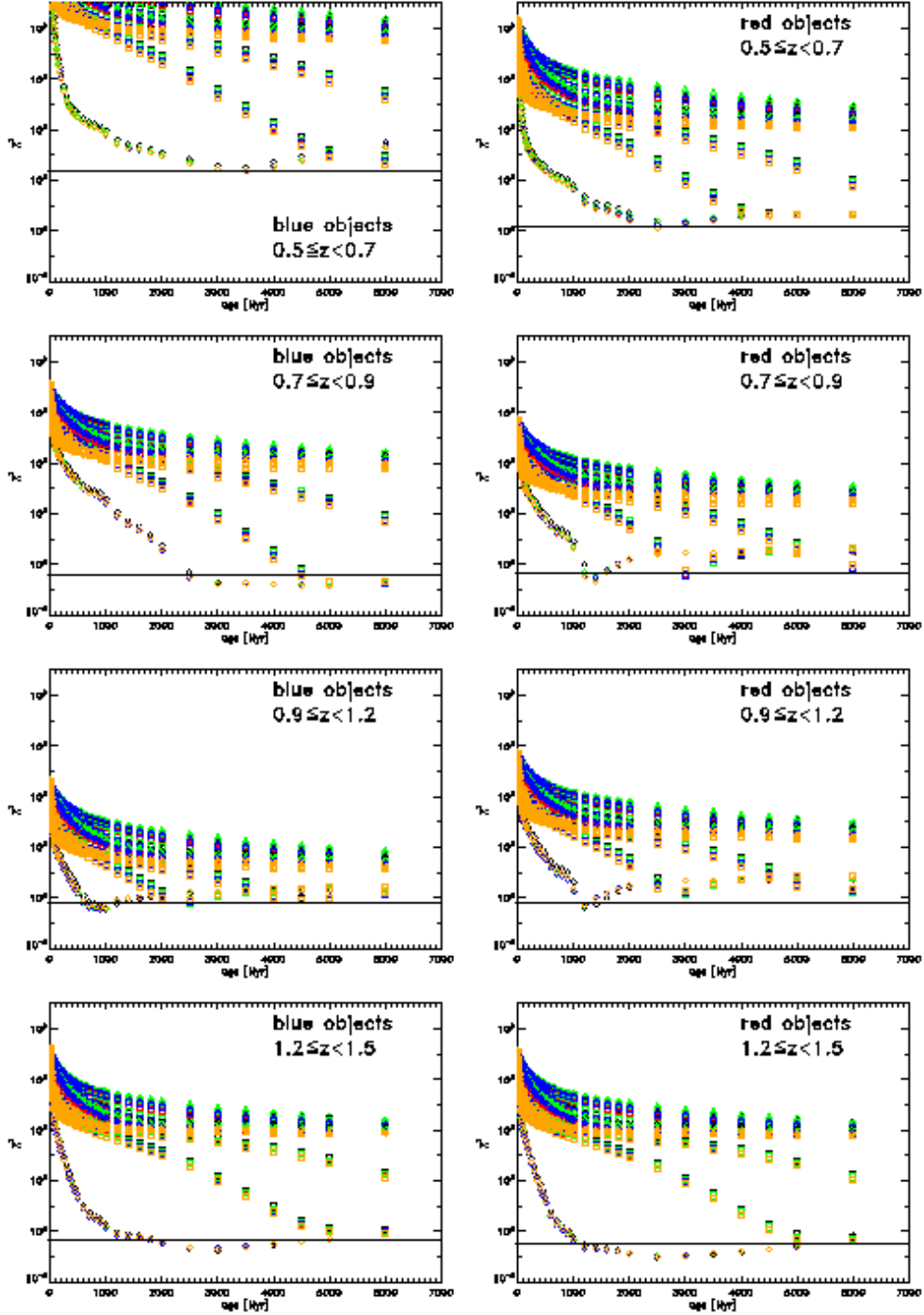}
\caption{\label{chi2_age_blue_red_ext}The same as
  Fig.~\ref{chi2_age_blue_red}, except for the inclusion of dust in the
  PEGASE models (see Fig.~\ref{lbg_blue_red_sfh_ext}). The notation of symbols
  and colors is identical to Fig.~\ref{chi2_age_ext}.}
\end{figure*}

\placefigure{lbg_blue_red_sfh}
\placefigure{chi2_age_blue_red}
\placefigure{lbg_blue_red_sfh_ext}
\placefigure{chi2_age_blue_red_ext}
The relatively large red to blue subsample sizes and the old ages for the LBG
candidate samples indicate that the populations are dominated by evolved
redder galaxies in comparison to the LBG sample of
\citet[][]{2007MNRAS.380..986B}. The best fitting model SEDs give mean
luminosity-weighted ages between 1.2 and 6~Gyr (see
Fig.~\ref{lbg_blue_red_sfh} and \ref{lbg_blue_red_sfh_ext}), with similar
results for blue and red subsamples. Dusty model estimates range over
1.0--5~Gyr.
The acceptable range of SED fits for the models including dust allows for
ages as young as 0.9~Gyr
(Figs.~\ref{chi2_age_blue_red},d).

\subsection{Luminosity Function}
We estimated the luminosity functions (LFs) for our LBG candidate samples
in the four different redshift bins using 1/V$_{max}$  
\citep[][]{1968ApJ...151..393S} 
We used k-corrected NUV magnitudes to estimate rest frame FUV fluxes, covered
by the GALEX NUV filter-band at our redshift intervals.  

For the 1/V$_{max}$ method, we followed the approach used by many other
studies \citep[e.g.][]{1993ApJ...404...51E,1995ApJ...455..108L,1996MNRAS.280..235E,2005ApJ...619L..43A,2006ApJ...647..853W} 
\begin{eqnarray}
\Phi(M,z) = \sum_{i=1}^{N}\frac{\omega_i}{V_{max,i}\Delta m}
\end{eqnarray}
where $\omega_i$ represents the weighting factor accounting for
incompleteness. The maximum volume a galaxy can be observed and still satisfy
the sample selection criteria is described by \citep[][]{1999astro.ph..5116H} 
\begin{eqnarray}
V_{max} = \left(\frac{c}{H_0}\right)^3d\Omega\left(\int_{z_{l}}^{z_{u}}{\frac{dz}{\sqrt{\Omega_m(1+z)^3+\Omega_{\Lambda}}}}\right)^3
\end{eqnarray}
For the integration limits we used the fixed limits $z_l$,$z_u$ of our four
different redshift intervals. We calculated the errors for the LF using
\begin{eqnarray}
\sigma_{\Phi}(M,z) =
\sqrt{\sum_{i=1}^{n}\left({\frac{\omega_i}{V_{max,i}\Delta m}}\right)^2}
\end{eqnarray} 
The results are compared to parameterizations of the Schechter function as
derived by \citet[][see Fig.~\ref{lbg_lfvmax}]{2005ApJ...619L..43A}. For the
FG1,FG2 and LQG0.8 samples we were able to cover the LF down to roughly
M$^{*}$. For the CCLQG at 1.2$\le z <$1.5 we were only able to derive the
LF for LBG candidates down to roughly 3M$^{*}$, including only two bright
magnitude bins. 

\placefigure{lbg_lfvmax}
The Schechter parameterization derived by
\citet[][]{2005ApJ...619L..43A} for all types of NUV-selected galaxies in
Fig.~\ref{lbg_lfvmax} is exceeded by the data, which implies that in all four
redshift bins the LBGs are over-abundant for their redshift.
In the two LQGs at $z\sim$0.8 and
$z\sim$1.3 the volume densities of the LBGs are consistent with
\citeauthor{2005ApJ...619L..43A} parameterization of
the LFs at 1.75$<z<$2.25 or LBGs at
2.5$<z<$3.5. In comparison to the two foreground samples (FG1 and FG2) we
also find an increase in the abundances of LBGs for the two LQGs. Although the
uncertainties in photometric redshifts are large and there is some
blending with the less dense
foreground regions, there is a clear indication for higher densities of star
forming galaxies in the two LQGs from the LFs. We have a sample
of 112 LBG candidates for the foreground LQG and 117 for the CCLQG. Since the
volume decreases by 43\% from $0.7\le z < 0.9$ to $1.2\le z < 1.5$, we
estimate an overdensity of 46$\pm$7\% or 2.6$\sigma$ for the CCLQG compared to
LQG0.8. The overall higher volume
densities for LBGs in all four redshift bins may at least partly be due to our
LBG selection criteria which includes galaxies which are relatively
evolved.

\subsection{LBG concentrations and LBG-quasar correlations} 

In Figure ~\ref{lbgconcen} we show the density maps of LBGs
photometrically-selected to be at $0.7<z<0.9$ (left panel) and $1.2<z<1.5$
(right panel). The LBG density maps were computed using a variant of the
adaptive kernel method \citep[][]{1986desd.book.....S} in which each LBG
in the redshift slice is represented by a Gaussian kernel
centered on the LBG. Following \citet[][]{2007MNRAS.381....7H}, we define the
width of the 
Gaussian kernel to be equal to the distance to the third nearest neighbor LBG
within the same redshift slice, and then calculate the local density at each
point as the sum of the Gaussian kernels. The isodensity contours in each plot
are linearly spaced at intervals of 20 LBGs deg$^{-2}$, the first contour
corresponding to an LBG density of 20 deg$^{-2}$. 

\placefigure{lbgconcen}
A number of distinct structures appear in the $z\sim 0.8$ LQG. For the CCLQG
at $z\sim$1.3 we were only able to detect the brightest LBG candidates with
 L$_{NUV}$$\ge$1.5$\times$10$^{12}$L$_\odot$. Several studies have
 indicated 
 that at $z<0.5-1$, quasars avoid both the highest density galaxy
 regions and 
 the field, instead preferentially populating cluster outskirts
 \citep[e.g.][]{2002MNRAS.331..569S,2004MNRAS.347.1241S,2008arXiv0804.1955K}.
 Such behavior is also suggested in Fig.~\ref{lbgconcen}.

We can test for quasar-LBG correlations in redshift slices using our photometric
redshift estimates, for comparison with higher redshift AGN-LBG correlations
from \citet[][]{2005ApJ...627L...1A}. The largest number of pairs would arise
for $0.7<z<0.9$, which contains 17 quasars and 117 LBGs. We calculated the
nearest neighbor distribution in angular distance in arcminute bins (1~arcmin is
0.8 co-moving Mpc), and compared it with 10000 randomly placed sets of
17 quasars 
within the GALEX fields. Results indicate mild ($\sim 2\sigma$) overdensities at
2--6~arcmin or $\sim 1.6-4.8$ co-moving Mpc (Fig.~\ref{lbgcorr}). This is
consistent with the LBG-AGN correlation length of
\citeauthor{2005ApJ...627L...1A}, and a factor of at least 3 smaller
than the overdensities around quasars measured by the proximity effect
at $2.1<z<3.3$ \citep[][]{2008MNRAS.389.1727D}.
Although it could be expected that $z\sim 1$
LBGs would be less massive than their $z\sim 3$ counterparts due to downsizing,
our sample appears to be dominated by massive galaxies and could well be similar
to the \citeauthor{2005ApJ...627L...1A} sample.  If quasars have similar
regions of enhanced density around them at $z\sim 0.8$ as at $2.1<z<3.3$,
then $z\sim 0.8$ LBGs would fall in regions of heightened density, but
not so high as quasars. 
\placefigure{lbgcorr}

\section{Summary and Conclusions}
\label{sumcon}
We present first results from the Clowes-Campusano LQG Survey, a 2 deg$^2$
multi-wavelength approach to study one of the largest structures with high
quasar density in
the high redshift Universe (z$\sim$1), the Clowes-Campusano Large Quasar
Group. The observations also covered a second LQG in front of the CCLQG at
z$\sim$0.8. Our data set includes GALEX FUV+NUV images covering a 2 deg$^2$
field and optical photometry of the NUV selected sample from the SDSS
DR5. With GALEX data, we reached a detection efficiency ranging between 80 --
90\,\%. The detection efficiency declines at m$_{FUV,NUV}\,=\,24$ due
to confusion and incompleteness. Using the FUV-dropout technique, selection
criteria adopted from \citet[][]{2006A&A...450...69B} and object
classification from SDSS DR5, we were able to select a sample of 1263 star
forming LBG candidates down to m$_{NUV} = 24.5$. Since photometric
redshift uncertainties increase significantly for galaxies with
m$_{NUV}\ge$23.5, we restricted further analysis to a subsample of 462
LBG candidates with m$_{NUV}\le$23.5.
We derived 7-band photometric redshifts
with accuracies $\sigma_{\Delta z}$\,=\,0.105 for all galaxies with $m_{NUV}
<23.5$ and $\sigma_{\Delta z_{LBG}}$\,=\,0.129 for the corresponding LBG candidates.
The
mean photometric redshift of the LBG candidate sample is
$\left<z\right>=0.86\pm 0.45$, and the majority of LBGs are at $z<1$, so we
mainly probe the foreground LQG. We derived star formation histories
for bright and faint LBG
candidate subsamples, and found relatively old best fitting
luminosity-weighted ages of 1.0-6 Gyr for models with and without
dust. Compared to the results of \citet[][]{2007MNRAS.380..986B}, who
estimated ages in the range of 250-500 Myr, our best fitting ages are
significantly older. This indicates that our sample is dominated by more
evolved, redder (and likely more massive)
LBG candidates. Dividing the LBG candidates into blue and red
subsamples using the mean stellar locus led to a similar conclusion.  
The best fitting SEDs for the blue and red subsamples yielded consistent
ages ranging between 1.0 and 6~Gyr.
Due to the high uncertainties of the SFHs resulting from the use of broad band
photometry and the large scatter of the averaged SEDs, it is important not to
over-interpret the results for the luminosity-weighted
ages. Our sample of LBG candidates includes only the most luminous
galaxies which formed stars over a long period, resulting in a significant
population of old red stars to counterbalance the young stars which produce
the Lyman break.
The faint LBGs in the LQG0.8 subsample show marginally larger
luminosity-weighted ages T$_L$.  A possible explanation is
lower
SFRs in the fainter subsample, which is supported by evidence for a
larger red and more evolved population of LBGs in LQG0.8 compared to
FG1 and FG2, which are not coincident with LQGs.

Possible effects of different environment densities can be more clearly
observed in the 
luminosity function for the four redshift slices. The LFs for LBG candidates
in the two LQGs show an increased volume density of star forming galaxies
compared to results of less dense regions in the CDF-S
\citep[][]{2005ApJ...619L..43A}. The LBG LF in the foreground LQG
(LQG0.8) is consistent with their 
parameterization of the Schechter function
corresponding to 1.75$<z<$2.25. 
Although we only have two
luminosity bins for the CCLQG, it also shows evidence for an overdensity
(more consistent with a population at 2.5$<z<$3.5).  Both redshift slices
containing LQGs have larger relative overdensities than the two redshift
slices which do not contain quasar overdensities.
We derived a sample of
112 LBG candidates for the foreground sample at 0.5$\le z<$0.7 (FG1) and 117
LBG candidates for the foreground LQG (LQG0.8) and although the volume
decreases by 43\% between those redshift intervals, the number of LBGs
stays about the same. This indicates an 
overdensity in star forming galaxies of 46$\pm$7\,\% or 2.6\,$\sigma$ compared
to the less dense foreground region FG1.
This leads to the conclusion that the high densities in both galaxies
\citep[e.g.][]{2002ApJ...578..708W,2004A&A...421..157H} and
QSOs is coincident with an overdensity of star forming galaxies due to LBGs and
QSOs; both  types of objects
trace an underlying overdensity of galaxies.

The LBGs in the LQG0.8 redshift slice appear to show substructure in
the host large quasar group, as shown in
the density plots in Fig.~\ref{lbgconcen}.
When compared to quasar locations in the same redshift range, the LBGs
and quasars show a marginal overdensity on angular scales corresponding to
1.6--4.8~ Mpc, such that quasars prefer the outskirts of
dense regions rather than the cores.  This result, if confirmed, would
be consistent with trends seen at $z<0.4$
\citep[][]{2002MNRAS.331..569S,2004MNRAS.347.1241S}, and also qualitatively
noted in a $z\sim 0.9$ supercluster
\citep[][]{2008arXiv0804.1955K,2008arXiv0809.2091K}.
It also is consistent with the view that gas-rich mergers cause quasar
activity, with such mergers preferentially occurring in regions with
excess small-scale galaxy overdensities but not in such dense regions that
gas stripping has largely taken place
\citep[][and references therein]{2008ApJS..175..356H}.

The two large quasar groups in the area surveyed here can provide
uniquely efficient sites for studying a wide variety of environments
and for quasar-galaxy relations.  Future studies will require
IR imagery to determine stellar masses of galaxies
and to refine photometric redshifts to $z\sim 1.5$, more spectra to
confirm the location and nature of clusters in the field and their
relation to the rich quasar environment, deeper UV observations to probe the
LBG luminosity function within the $1.2<z<1.5$ Clowes-Campusano LQG.
  X-ray observations would be able to confirm virialized regions within the
  LQGs.

\acknowledgments

    This publication makes use of data from GALEX Cycle 1 observations and
    is supported by NASA grant NNX05GK42G. Chris Hains acknowledges financial
    support from STFC. We thank David Koo, David Rosario
    and Dave Schiminovich for their very useful comments. We also thank the
    referee for her/his comments which helped to improve the paper
    significantly.

    LEC received partial support from Center of Excellence in
    Astrophysics and Associated Technologies (PFB 06).

    Funding for the Sloan Digital Sky Survey (SDSS) and SDSS-II has been
    provided by the Alfred P. Sloan Foundation, the Participating
    Institutions, the National Science Foundation, the U.S. Department of
    Energy, the National Aeronautics and Space Administration, the Japanese
    Monbukagakusho, and the Max Planck Society, and the Higher Education
    Funding Council for England. The SDSS Web site is http://www.sdss.org/. 

    The SDSS is managed by the Astrophysical Research Consortium (ARC) for the
    Participating Institutions. The Participating Institutions are the
    American Museum of Natural History, Astrophysical Institute Potsdam,
    University of Basel, University of Cambridge, Case Western Reserve
    University, The University of Chicago, Drexel University, Fermilab, the
    Institute for Advanced Study, the Japan Participation Group, The Johns
    Hopkins University, the Joint Institute for Nuclear Astrophysics, the
    Kavli Institute for Particle Astrophysics and Cosmology, the Korean
    Scientist Group, the Chinese Academy of Sciences (LAMOST), Los Alamos
    National Laboratory, the Max-Planck-Institute for Astronomy (MPIA), the
    Max-Planck-Institute for Astrophysics (MPA), New Mexico State University,
    Ohio State University, University of Pittsburgh, University of Portsmouth,
    Princeton University, the United States Naval Observatory, and the
    University of Washington.



Facilities: \facility{GALEX}, 
\facility{Sloan}, 
\facility{Magellan:Baade(IMACS)}



\appendix





\clearpage
\bibliographystyle{aa}
\bibliography{references_lqg}

\clearpage




\clearpage

\clearpage

\clearpage

\clearpage

\clearpage

\clearpage

\clearpage

\clearpage

\clearpage

\clearpage

\clearpage

\clearpage

\clearpage

\clearpage

\clearpage

\clearpage

\clearpage

\clearpage
\setcounter{figure}{11}
\begin{figure*}
\epsscale{1.0}
\plotone{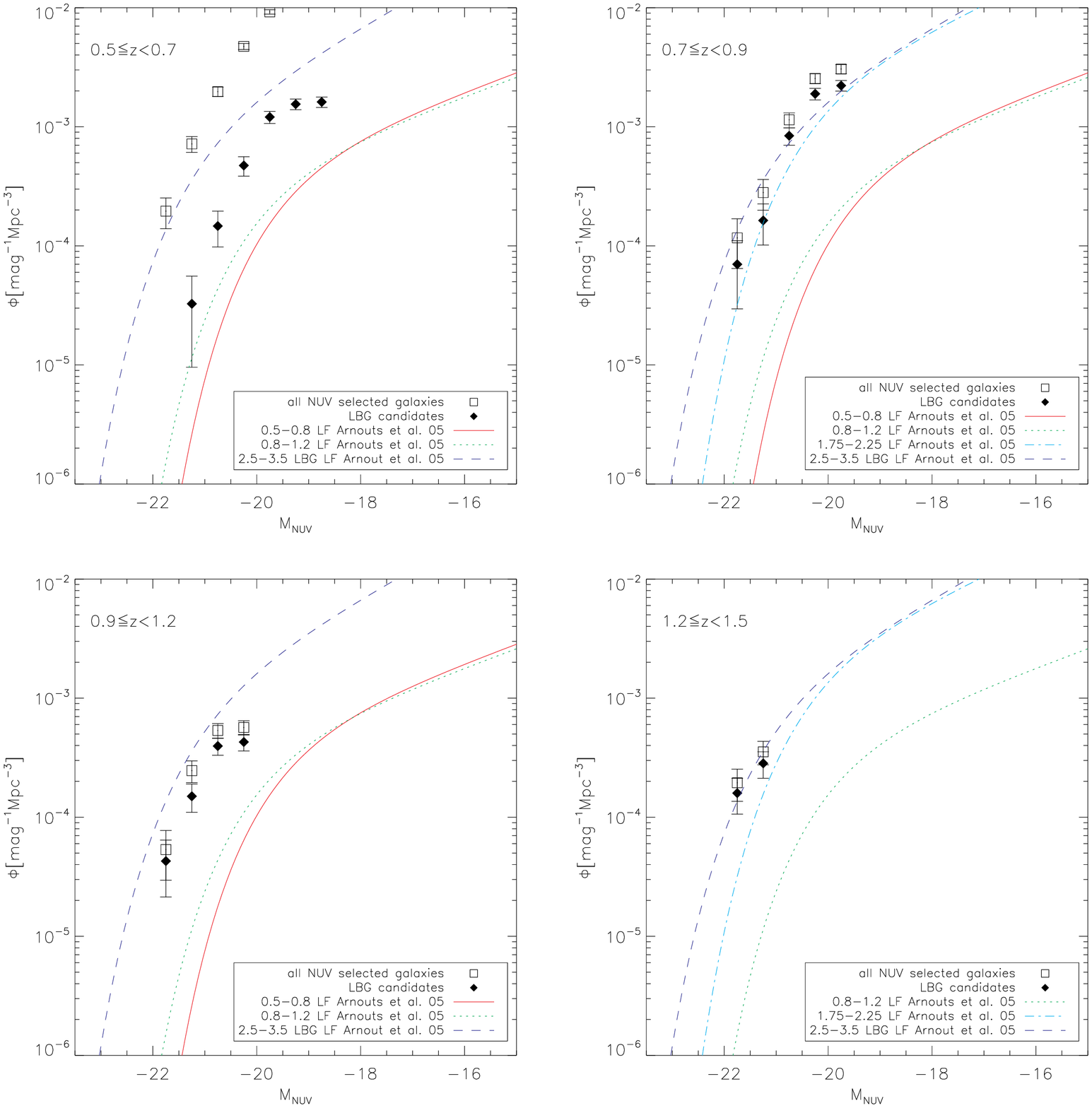}
\caption{\label{lbg_lfvmax} Luminosity Functions (LFs) for the LBG candidate
  samples (filled diamonds) and the all NUV selected galaxies (open squares) in
  the four different redshift bins. The LFs were derived using the
  1/V$_{max}$-method. For comparison we plotted LFs for different redshift
  intervals and selection criteria of \citet[][]{2005ApJ...619L..43A}.}
\end{figure*}

\clearpage
\begin{figure*}
\epsscale{0.8}
\plottwo{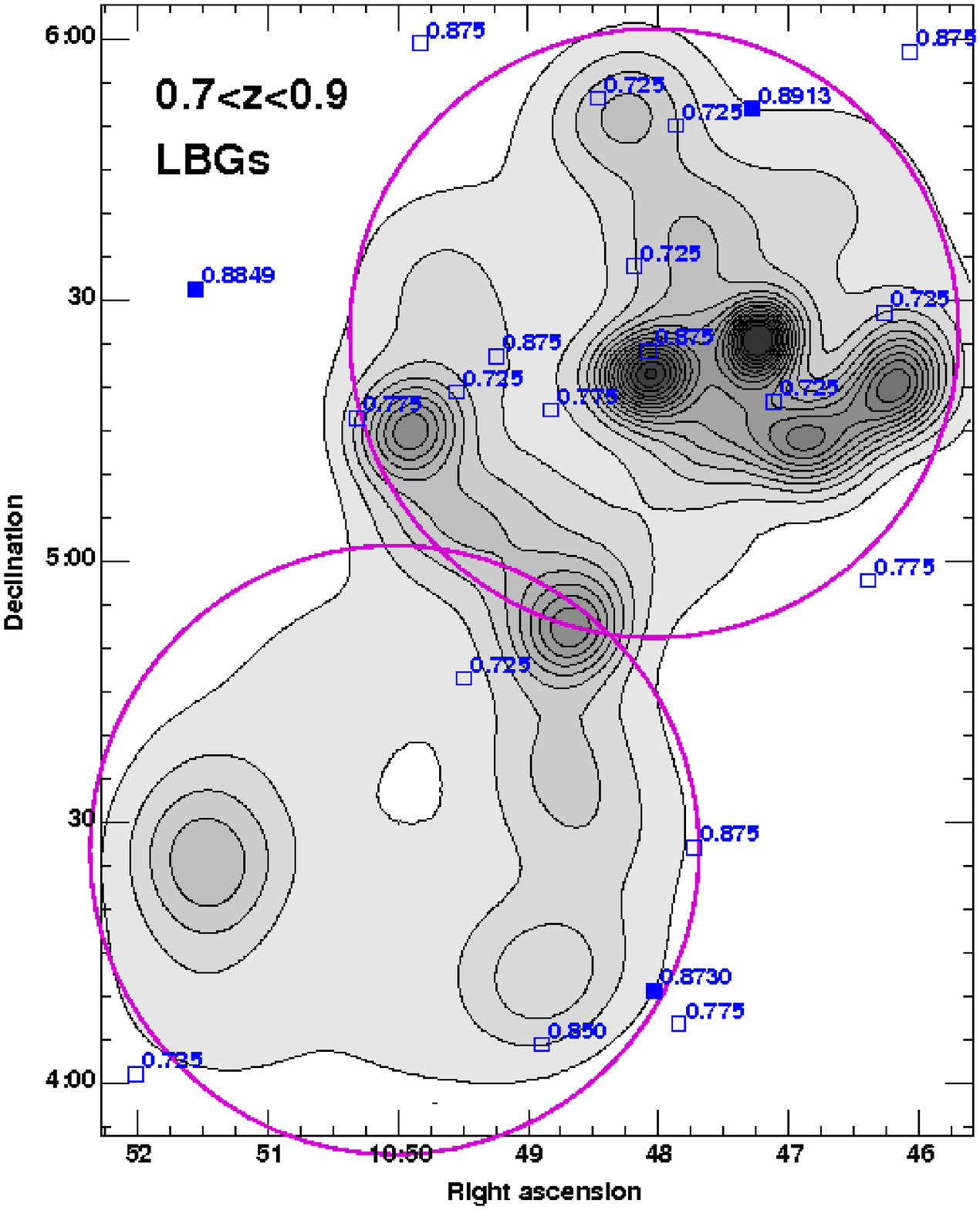}{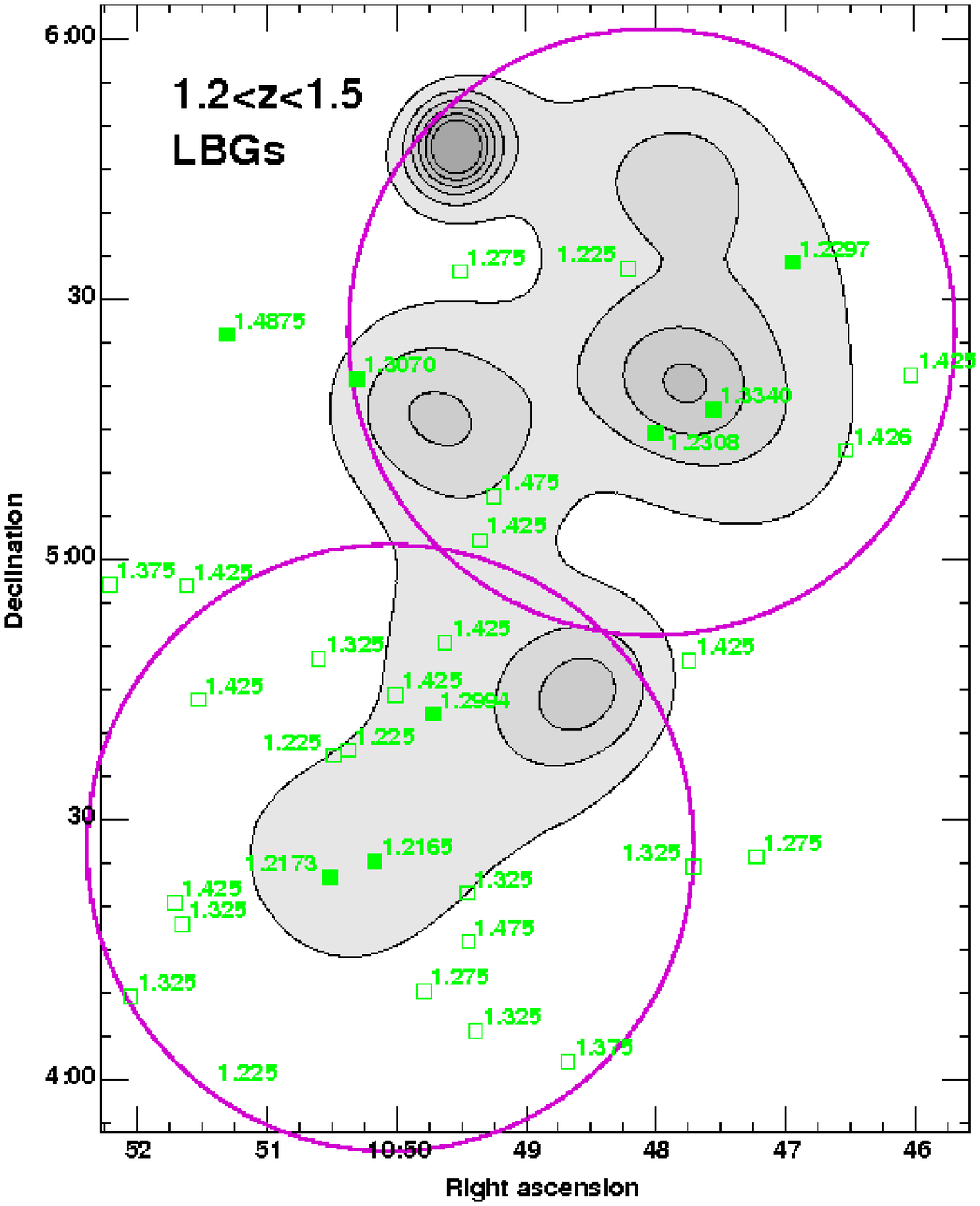}
\caption{Density plot for the LBGs in the two LQGs at $z\sim$0.8 (left with
  blue/dark quasar labels) and $z\sim$1.3 (right with green/light quasar
  labels). Filled squares represent spectroscopic and open photometric quasar
  samples. The two GALEX pointings are indicated by the large 
  circles. The lowest density contour represents a density of 20 LBGs
  deg$^{-2}$ with a separation of 20 LBGs deg$^{-2}$ between
  contours. Both LQGs show concentrations and structures of LBGs (center of
  the northern and southeast in the southern GALEX field).}
\label{lbgconcen}
\end{figure*}

\clearpage
\begin{figure}
\epsscale{0.8}
\plotone{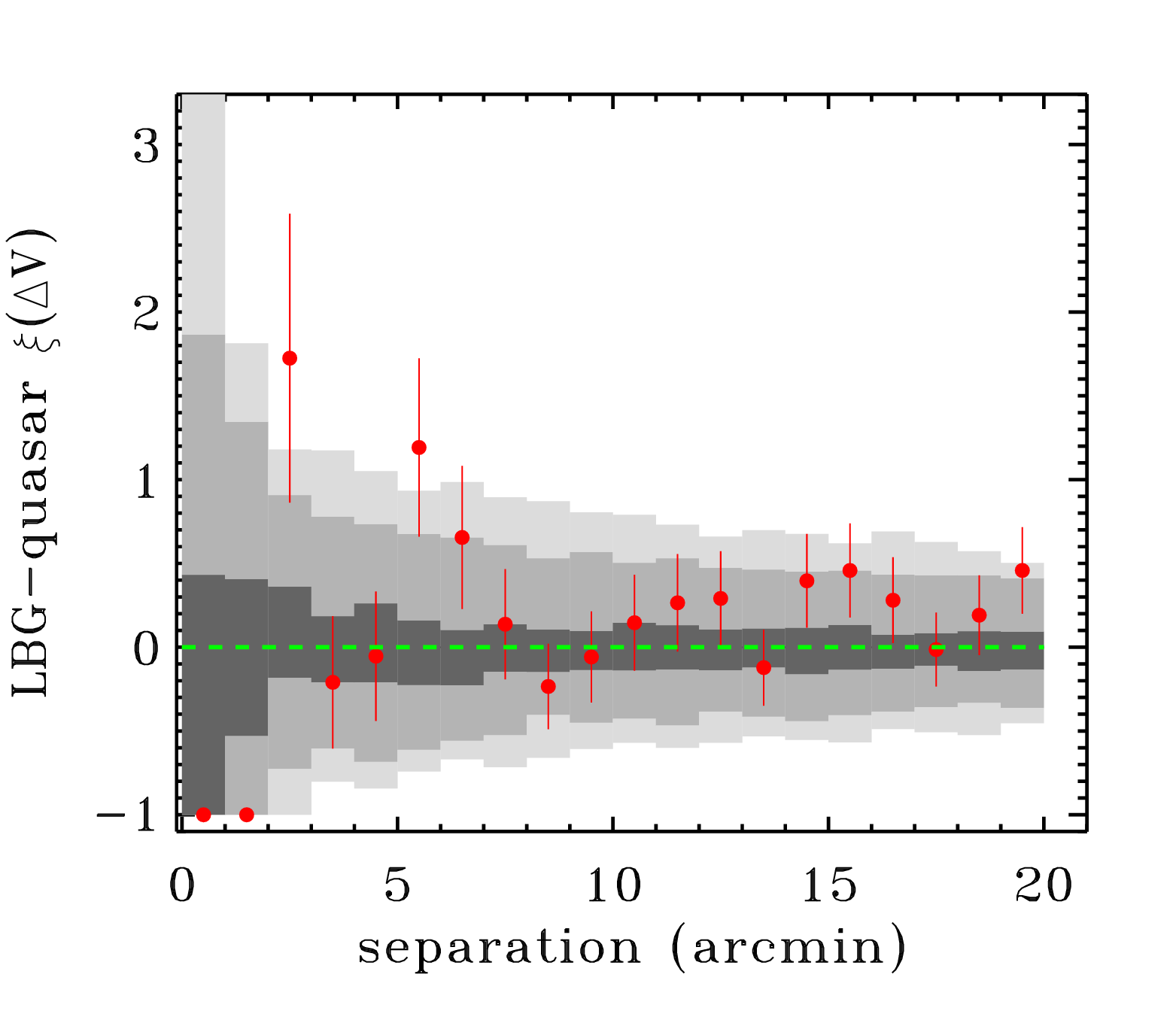}
\caption{Two point correlation function for LBGs (117) and quasars (17) at
  $0.7<z<0.9$, within the two GALEX fields.  This redshift slice provides the
  largest combination of LBG-quasar pairs. Shaded regions denote 68, 95, 99\%
  confidence limits from 10,000 Monte Carlo simulations of sets of 17
  random quasar RA and DEC placements in the observed area. Error
  bars show example 
  $1\sigma$ Poissonian errors. At $z=0.8$, 1 arcmin corresponds to 0.45
  local-frame or 0.81 co-moving Mpc. There are no LBG-quasar pairs
  within 2 arcmin of each other. The signal at 2-3 arcmin arises from 10
  pairs. None of the 10,000 simulations had more than 9 in that bin, and
  we estimate the significance at $3.3\sigma$. The signal at 5-6 arcmin arises
  from 17 pairs. None of the 10,000 simulations had more than 15 in that
  bin, and we estimate the significance at $3.1\sigma$. However, the $\sim
  1.8\sigma$ overdensity at 6-7 arcmin separation, and the lack of an
    overdensity at 3-4~arcmin, implies that the overdensity at
  separation 5-7 arcmin is more robust. 
}
\label{lbgcorr}
\end{figure}

\clearpage
\begin{deluxetable}{l c c c c l}
\tabletypesize{\scriptsize}
\tablewidth{5.1in}
\tablecaption{\label{exposuretimes} Observation summary}
\tablehead{
\colhead{}&
\multicolumn{2}{c}{FUV}&
\multicolumn{2}{c}{NUV}&
\colhead{}\\
\colhead{Field ID}&
\colhead{T$\rm_{exp}$}&
\colhead{m$\rm_{lim}$}&
\colhead{T$\rm_{exp}$}&
\colhead{m$\rm_{lim}$}&
\colhead{Note}\\
\colhead{}&
\colhead{[sec]}&
\colhead{[mag]}&
\colhead{[sec]}&
\colhead{[mag]}&
\colhead{}\\
\colhead{(1)}&
\colhead{(2)}&
\colhead{(3)}&
\colhead{(4)}&
\colhead{(5)}&
\colhead{(6)}
}
\startdata
21240-GI1\_035001\_J104802p052610&22902&24&38624&24.0&northern GALEX field\\
21241-GI1\_035002\_J105002p042644&20817&24&33021&24.0&southern GALEX field\\
\enddata
\tablecomments{The total magnitudes in columns (3) and (5) are the 80~\%
  completeness limits.}
\end{deluxetable}

\clearpage
\begin{deluxetable}{@{}c@{}@{}c@{}@{}c@{}@{}c@{}@{}c@{}@{}c@{}@{}c@{}@{}c@{}@{}c@{}@{}c@{}@{}c@{}@{}c@{}@{}c@{}@{}c@{}@{}c@{}@{}c@{}@{}c@{}@{}c@{}@{}c@{}@{}c@{}} 
\tabletypesize{\scriptsize}
\rotate
\tablewidth{22.7cm}
\tablecaption{\label{lbg_example} Examples for LBG candidates}
\tablehead{
\colhead{Id}&
\colhead{RA(J2000)}&
\colhead{DEC(J2000)}&
\colhead{FUV}&
\colhead{\D FUV}&
\colhead{NUV}&
\colhead{\D NUV}&
\colhead{u}&
\colhead{\D u}&
\colhead{g}&
\colhead{\D g}&
\colhead{r}&
\colhead{\D r}&
\colhead{i}&
\colhead{\D i}&
\colhead{z}&
\colhead{\D z}&
\colhead{$z$}&
\colhead{\D$z$}\\
\colhead{(1)}&
\colhead{(2)}&
\colhead{(3)}&
\colhead{(4)}&
\colhead{(5)}&
\colhead{(6)}&
\colhead{(7)}&
\colhead{(8)}&
\colhead{(9)}&
\colhead{(10)}&
\colhead{(11)}&
\colhead{(12)}&
\colhead{(13)}&
\colhead{(14)}&
\colhead{(15)}&
\colhead{(16)}&
\colhead{(17)}&
\colhead{(18)}&
\colhead{(19)}
}
\startdata
LQG\_J104745+45136&10:47:45.93&4:51:36.82&25.601&0.855&23.391&0.035&21.880&0.240&21.409&0.061&20.294&0.032&19.807&0.031&19.520&0.104&0.576&0.034\\
LQG\_J104832+45217&10:48:32.36&4:52:17.21&25.732&1.173&23.064&0.032&22.197&0.330&22.282&0.137&21.765&0.126&21.230&0.114&20.961&0.397&0.663&0.220\\
LQG\_J104808+45223&10:48:08.49&4:52:23.37&99.000&99.000&23.314&0.033&22.776&0.446&22.416&0.123&22.119&0.128&21.479&0.106&20.846&0.269&1.154&0.285\\
LQG\_J104710+45329&10:47:10.68&4:53:29.16&25.419&0.779&23.382&0.037&22.420&0.287&22.588&0.121&22.060&0.102&21.911&0.125&21.632&0.428&0.348&0.262\\
LQG\_J104726+45345&10:47:26.29&4:53:45.69&23.411&0.249&20.791&0.007&18.902&0.021&17.756&0.006&17.321&0.005&17.265&0.006&17.222&0.014&0.098&0.015\\
LQG\_J104850+45451&10:48:50.52&4:54:51.50&22.527&0.132&19.450&0.002&15.523&0.009&15.270&0.012&14.503&0.010&11.138&0.001&11.324&0.002&0.851&0.008\\
LQG\_J104656+45529&10:46:56.13&4:55:29.04&22.913&0.184&18.293&0.001&15.433&0.010&11.975&0.001&14.659&0.011&11.102&0.000&13.599&0.016&1.196&0.047\\
LQG\_J104718+45617&10:47:18.30&4:56:17.15&24.271&0.390&20.171&0.003&16.387&0.006&15.045&0.003&14.554&0.004&17.364&0.017&14.330&0.004&1.929&0.004\\
LQG\_J104829+45618&10:48:29.77&4:56:18.67&31.161&180.875&23.206&0.037&22.094&0.782&22.418&0.398&21.160&0.191&19.959&0.096&19.977&0.443&0.798&0.098\\
LQG\_J104735+45712&10:47:35.17&4:57:12.04&25.674&0.956&23.360&0.036&24.247&1.215&22.643&0.159&22.091&0.134&21.638&0.134&20.858&0.289&1.021&0.427\\
LQG\_J104754+45745&10:47:54.93&4:57:45.61&23.499&0.211&20.416&0.004&16.573&0.006&15.109&0.004&14.609&0.004&14.882&0.003&14.420&0.005&0.055&0.032\\
LQG\_J104754+45804&10:47:54.08&4:58:04.16&24.854&0.534&22.771&0.025&22.448&0.428&21.809&0.093&20.712&0.052&20.413&0.057&20.380&0.241&0.416&0.090\\
LQG\_J104851+45808&10:48:51.81&4:58:08.03&24.400&0.498&20.785&0.006&16.396&0.006&14.694&0.004&14.067&0.004&14.119&0.001&13.726&0.004&0.050&0.029\\
LQG\_J104738+45837&10:47:38.29&4:58:37.40&25.209&0.710&23.183&0.034&24.097&1.460&22.021&0.113&21.498&0.109&20.933&0.091&20.573&0.297&0.647&0.106\\
LQG\_J104839+45850&10:48:39.31&4:58:50.02&25.045&0.519&22.835&0.021&22.360&0.282&22.253&0.097&21.842&0.094&21.614&0.112&21.025&0.285&0.553&0.198\\
\enddata
\tablecomments{Summary of LBG properties. column (2): RA in hh:mm:ss.ss;
  column (3) DEC in +dd:mm:ss.ss; column (4) - (17): magnitudes and
  errors of our 7 band photometry; column (18)+(19): photometric redshifts and
  errors derived using Hyperz\\  
  The complete version of this table is in the electronic edition
  of the Journal. The printed edition contains only a sample.}
\end{deluxetable}

\clearpage
\begin{deluxetable}{c c c c c c c c c c c c c c}
\tabletypesize{\scriptsize}
\rotate
\tablewidth{22cm}
\tablecaption{\label{subsample} Summary of selected subsamples}
\tablehead{
\colhead{name}&
\colhead{$z$}&
\colhead{size}&
\colhead{FUV}&
\colhead{NUV}&
\colhead{u}&
\colhead{g}&
\colhead{r}&
\colhead{i}&
\colhead{z}&
\colhead{FUV-NUV}&
\colhead{u-g}&
\colhead{g-r}&
\colhead{i-z}\\
\colhead{(1)}&
\colhead{(2)}&
\colhead{(3)}&
\colhead{(4)}&
\colhead{(5)}&
\colhead{(6)}&
\colhead{(7)}&
\colhead{(8)}&
\colhead{(9)}&
\colhead{(10)}&
\colhead{(11)}&
\colhead{(12)}&
\colhead{(13)}&
\colhead{(14)}

}
\startdata
FG1&0.5$\le$z$<$0.7&64&25.72$\pm$0.60&23.22$\pm$0.23&22.89$\pm$0.52&22.28$\pm$0.35&21.54$\pm$0.48&21.07$\pm$0.58&20.84$\pm$0.68&2.50&0.61&0.74&0.23\\
LQG0.8&0.7$\le$z$<$0.9&73&25.97$\pm$1.07&23.08$\pm$0.29&22.69$\pm$0.60&22.27$\pm$0.45&21.77$\pm$0.59&21.05$\pm$0.55&20.90$\pm$0.76&2.90&0.42&0.50&0.15\\
FG2&0.9$\le$z$<$1.2&35&25.94$\pm$0.78&23.08$\pm$0.42&22.40$\pm$0.68&22.04$\pm$0.47&21.80$\pm$0.51&21.45$\pm$0.57&21.08$\pm$0.50&2.86&0.35&0.25&0.38\\
CCLQG&1.2$\le$z$<$1.5&25&26.16$\pm$1.75&22.07$\pm$1.27&19.48$\pm$2.70&18.86$\pm$3.25&18.72$\pm$3.34&18.54$\pm$3.25&17.83$\pm$3.42&4.08&0.63&0.14&0.72\\
\enddata
\tablecomments{The subsamples are selected for the two foreground regions
  (FG1+FG2) and in the LQGs (LQG0.8+CCLQG). Column (2) gives the
  redshift interval over which the subsamples were averaged. Columns (4)-(10)
  present the averaged total magnitudes in the FUV+NUV and 5 SDSS filter bands
  and columns (11)-(14) summarize the averaged colors for the four subsamples.}
\end{deluxetable}
 
\clearpage
\begin{deluxetable}{l c c l c l c l}
\tablecaption{\label{sedfitres_noext} SED fitting results for bright-faint
  subsamples} 
\tablehead{
\colhead{name}&
\colhead{redshift}&
\multicolumn{2}{c}{bright    }&
\multicolumn{2}{c}{faint     }&
\multicolumn{2}{c}{all       }\\
\colhead{}&
\colhead{z}&
\colhead{best}&
\colhead{SFLaw}&
\colhead{best}&
\colhead{SFLaw}&
\colhead{best}&
\colhead{SFLaw}\\
\colhead{}&
\colhead{}&
\colhead{[Gyr]}&
\colhead{}&
\colhead{[Gyr]}&
\colhead{}&
\colhead{[Gyr]}&
\colhead{}\\
\colhead{(1)}&
\colhead{(2)}&
\colhead{(3)}&
\colhead{(4)}&
\colhead{(5)}&
\colhead{(6)}&
\colhead{(7)}&
\colhead{(8)}
}
\startdata
FG1&0.5$\le$z$<$0.7&6.0&expo. decr.&6.0&constant&6.0&expo. decr.\\
LGQ0.8&0.7$\le$z$<$0.9&3.5&expo. decr.&6.0&constant&3.5&expo. decr.\\
FG2&0.9$\le$z$<$1.2&6.0&constant&\nodata&\nodata&6.0&constant\\
CCLQG&1.2$\le$z$<$1.5&6.0&constant&\nodata&\nodata&6.0&constant\\
\enddata
\tablecomments{Results for the SED fits of the LBG subsamples in the individual
  redshift bins using PEGASE models without extinction by dust. The table is
  split into three blocks for the results of the bright (columns (3)+(4)), faint
  (columns (5)+(6)) LBG subsample as well as the results for the complete LBG
  candidate  sample (columns (7)+(8)). Every block consists of one column for
  the age and the SFLaw of the best fitting model. The redshift bin is indicated
  in column (2).
}
\end{deluxetable}

\clearpage
\begin{deluxetable}{@{}l@{ } @{}c@{} @{}c@{} @{}l@{ } @{}l@{} @{}c@{} @{}l@{ } @{}l@{} @{}c@{} @{}l@{} @{}l@{}}
\tabletypesize{\scriptsize}
\tablewidth{6.7in}
\tablecaption{\label{sedfitres_ext} SED fitting results for bright-faint
  subsamples including extinction}
\tablehead{
\colhead{name}&
\colhead{redshift}&
\multicolumn{3}{c}{bright    }&
\multicolumn{3}{c}{faint     }&
\multicolumn{3}{c}{all       }\\
\colhead{}&
\colhead{z}&
\colhead{best}&
\colhead{SFLaw}&
\colhead{dust geometry}&
\colhead{best}&
\colhead{SFLaw}&
\colhead{dust geometry}&
\colhead{best}&
\colhead{SFLaw}&
\colhead{dust geometry}\\
\colhead{}&
\colhead{}&
\colhead{[Gyr]}&
\colhead{}&
\colhead{[Gyr]}&
\colhead{}&
\colhead{[Gyr]}&
\colhead{}\\
\colhead{(1)}&
\colhead{(2)}&
\colhead{(3)}&
\colhead{(4)}&
\colhead{(5)}&
\colhead{(6)}&
\colhead{(7)}&
\colhead{(8)}&
\colhead{(9)}&
\colhead{(10)}&
\colhead{(11)}
}
\startdata
FG1&0.5$\le$z$<$0.7&4.0&expo. decr.&face-on disk&2.5&burst&spherical&2.5&burst&edge-on disk\\
LGQ0.8&0.7$\le$z$<$0.9&1.4&burst&edge-on disk&3.5&expo. decr.&45$^\circ$
incl. disk&1.4&burst&edge-on disk\\
FG2&0.9$\le$z$<$1.2&1.0&burst&face-on disk&\nodata&\nodata&\nodata&1.0&burst&face-on disk\\
CCLQG&1.2$\le$z$<$1.5&3.0&burst&spherical&\nodata&\nodata&\nodata&3.0&burst&spherical\\
\enddata
\tablecomments{Results for the SED fitting of the LBG subsamples for the
  individual redshift bins as described in Table~\ref{sedfitres_noext}
  except now including extinction due to dust in the
  PEGASE models.}
\end{deluxetable}

\clearpage
\begin{deluxetable}{l c c c c r}
\tablewidth{4.5in}
\tablecaption{\label{sedfitres_color} SED fitting results for color selected
  LBG subsamples}
\tablehead{
\colhead{name}&
\colhead{redshift}&
\multicolumn{2}{c}{blue    }&
\multicolumn{2}{c}{red     }\\
\colhead{}&
\colhead{z}&
\colhead{best}&
\colhead{SFLaw}&
\colhead{best}&
\colhead{SFLaw}\\
\colhead{}&
\colhead{}&
\colhead{[Gyr]}&
\colhead{}&
\colhead{[Gyr]}&
\colhead{}\\
\colhead{(1)}&
\colhead{(2)}&
\colhead{(3)}&
\colhead{(4)}&
\colhead{(5)}&
\colhead{(6)}\\
}
\startdata
FG1&0.5$\le$z$<$0.7&6.0&constant&6.0&constant\\
LGQ0.8&0.7$\le$z$<$0.9&6.0&constant&3.5&expo. decr.\\
FG2&0.9$\le$z$<$1.2&6.0&constant&6.0&constant\\
CCLQG&1.2$\le$z$<$1.5&6.0&constant&6.0&constant\\
\enddata
\tablecomments{Results for the SED fitting of the
  LBG color selected subsamples using models without extinction by dust. The
  selection was done according to their location in the g-i vs. r-i
  color-color diagram. The MSL has been used to separate the blue and red
  subsample. The table is split into two blocks for the results of the blue
  (columns (3)+(4)), and red (columns (5)+(6)) LBG subsample. Every block
  consists of one column for the age and the SFLaw of the best fitting
  model. The redshift bin is indicated in column (2).}
\end{deluxetable}

\clearpage
\begin{deluxetable}{l c c l l c l l}
\tablecaption{\label{sedfitresext_color} SED fitting results for color
  selected LBG subsamples including extinction}
\tablehead{
\colhead{name}&
\colhead{redshift}&
\multicolumn{3}{c}{blue    }&
\multicolumn{3}{c}{red     }\\
\colhead{}&
\colhead{z}&
\colhead{best}&
\colhead{SFLaw}&
\colhead{dust geometry}&
\colhead{best}&
\colhead{SFLaw}&
\colhead{dust geometry}\\
\colhead{}&
\colhead{}&
\colhead{[Gyr]}&
\colhead{}&
\colhead{}&
\colhead{[Gyr]}&
\colhead{}&
\colhead{}\\
\colhead{(1)}&
\colhead{(2)}&
\colhead{(3)}&
\colhead{(4)}&
\colhead{(5)}&
\colhead{(6)}&
\colhead{(7)}&
\colhead{(8)}\\
}
\startdata
FG1&0.5$\le$z$<$0.7&3.5&burst&face-on disk&2.5&burst&face-on disk\\
LGQ0.8&0.7$\le$z$<$0.9&5.0&burst&face-on disk&1.4&burst&face-on\\
FG2&0.9$\le$z$<$1.2&1.0&burst&spherical&1.2&burst&spherical\\
CCLQG&1.2$\le$z$<$1.5&3.0&burst&spherical&2.5&burst&spherical\\
\enddata
\tablecomments{Results for the SED fitting of the LBG
  subsamples as described in Table~\ref{sedfitres_color} now including
  extinction due to dust in the model SEDs.}
\end{deluxetable}









\end{document}